\documentclass[]{ptephy_v1}

\usepackage{amsmath}
\usepackage{url}

\usepackage{graphicx}
\usepackage{color}

\begin{document}

\title{Can an off-axis gamma-ray burst jet in GW170817 explain all the electromagnetic counterparts?}

\author{\name{Kunihito Ioka}{1} and \name{Takashi Nakamura}{1,2}}

\address{${}^1$\affil{1}{Center for Gravitational Physics, Yukawa Institute for Theoretical Physics, Kyoto University, Kyoto 606-8502, Japan}
\\
${}^2$\affil{2}{Department of Physics, Kyoto University, Kyoto 606-8502, Japan}}

\begin{abstract}
Gravitational waves from a merger of two neutron stars (NSs) were discovered
for the first time in GW170817, 
together with diverse electromagnetic (EM) counterparts.
To make constraints on a relativistic jet from the NS merger,
we calculate the EM signals
in (1) the short gamma-ray burst sGRB 170817A from an off-axis jet,
(2) the optical--infrared macronova (or kilonova), especially the blue macronova, 
from a jet-powered cocoon,
and (3) the X-ray and radio afterglows
from the interaction between the jet and interstellar medium.
We find that a typical sGRB jet is consistent with these observations,
and there is a parameter space to explain all the observations
in a unified fashion
with an isotropic energy $\sim 10^{51}$--$10^{52}$ erg,
opening angle $\sim 20^{\circ}$,
and viewing angle $\sim 30^{\circ}$.
The off-axis emission is less de-beamed than the point-source case
because the viewing angle is comparable to the opening angle.
We also analytically show that the jet energy accelerates
a fair fraction of the merger ejecta
to a sub-relativistic velocity $\sim 0.3$--$0.4$c
as a cocoon in a wide parameter range.
The ambient density might be low $\sim 10^{-3}$--$10^{-6}$ cm$^{-3}$,
which can be tested by future observations of 
radio flares and X-ray remnants.
\end{abstract}

\subjectindex{E01, E02, E32, E35, E37}

\maketitle

\section{Introduction}

At last, gravitational wave (GW) astronomy has truly started
with the discovery of GWs from a merger of two neutron stars (NSs), called GW170817,
by the Laser Interferometer Gravitational-Wave Observatory (LIGO) 
and the Virgo Consortium (LVC) \citep{GW170817}
and the follow-up discoveries of electromagnetic (EM) counterparts \citep{GW-EM17}.
This historical milestone comes a century after Einstein predicted the existence of GWs,\footnote{
The announcement was made two months after the rumors spread \citep{Castelvecchi17}.}
$30$--$40$ years after the indirect discoveries of GWs \citep{Hulse_Taylor75,Taylor_Weisberg89},
and two years after the direct discoveries of GWs from black hole (BH) mergers
\citep{GW150914,GW151226,GW170104,GW170814},
for which the Nobel Prize in Physics 2017 was awarded.
For the GWs from BH mergers,
no EM counterparts have been detected despite intensive efforts
(see, e.g., Refs.~\citep[][]{GW150914-localization,Evans+16,Morokuma+16,Adriani+16,Yoshida+17}),
as expected from the theoretical grounds (see, e.g., Refs.~\citep[][]{Woosley+16,Ioka+17,Kimura+17,Fedrow+17}),
except for a claim for detection with GW150914 by
the Gamma-Ray Burst Monitor (GBM) on the {\it Fermi} satellite
({\it Fermi}/GBM) \citep{Connaughton+16},
which is questioned by the INTEGRAL group \citep{Savchenko+16} 
and the GBM team members \citep{Greiner+16}.
Because of the poor sky localization with GWs,
even a host galaxy has not been identified so far.
In GW170817, the situation has been revolutionized by the discovery of EM counterparts.

Two seconds ($\sim 1.7\,{\rm s}$) after GW170817,
{\it Fermi}/GBM was triggered by a short (duration $\sim 2$ s) gamma-ray burst (sGRB)
consistent with the GW localization, called sGRB 170817A \citep{GW-GRB17,Fermi/GBM17}.
INTEGRAL also detected a similar $\gamma$-ray flux with $\sim 3 \sigma$ \citep{GW-GRB17,Integral17}.
This was followed by ultraviolet, optical, and infrared detections \citep{GW-EM17,Coulter+17,Tanaka+17b,Utsumi+17,Tominaga+17,Drout+17,Swift-NuSTAR17,Arcavi+17,Smartt+17,Shappee+17,Pian+17,Kasen+17,Kasliwal+17b,Tanvir+17,Kilpatrick+17,Soares-Santos+17,Cowperthwaite+17,Nicholl+17,Chornock+17,Valenti+17,Diaz+17,McCully+17,Buckley+17}.
In addition, X-ray and radio afterglows were also discovered \citep{Chandra17,Chandra17b,Chandra17c,Hallinan+17,Alexander+17}.
These EM observations find the host galaxy NGC 4993 at a distance of $\approx 40$ Mpc \citep{GW-EM17}.
Remarkably, the world-wide follow-ups involve more than 3000 people \citep{GW-EM17}.

EM counterparts associated with binary NS mergers have long been considered and anticipated
(see, e.g., Refs.~\citep[][]{Metzger_Berger12,Rosswog15,Fernandez_Metzger16,Tanaka16}):
\begin{enumerate}
\item First, a binary NS merger is a promising candidate for the origin of sGRBs
\citep{Paczynski86,Goodman+86,Eichler+89}.
An sGRB is one of the brightest EM events in the Universe, caused by a relativistic jet.
A typical sGRB within the current GW horizon $\sim 100$ Mpc 
should be very bright if the jet points to us,
while an off-axis jet is generally very faint \citep{KI_Nakamura01,Yamazaki+02}
and hence an sGRB is not seriously thought to be the first to be detected,
considering a low probability for an on-axis jet at first glance.

\item Second, an sGRB jet produces an afterglow in broad bands 
via interaction with the interstellar medium (ISM) \citep{Sari+98}.
For off-axis observers, the early afterglow looks faint \citep{Granot+02},
and the decaying nature of the afterglow emission makes the detection not so easy.

\item Third, a small amount of NS material ejected from the NS mergers 
is expected to emit optical--infrared signals \citep{Li_Paczynski98},
the so-called ``macronova'' \citep{Kulkarni05} and ``kilonova'' \citep{Metzger+10}.\footnote{
We use ``macronova'' as it was invented earlier than ``kilonova''.
In addition, as we discuss, there could be other energy sources than the $r$-process elements
and the energy source cannot specify the name, as in the case of ``supernova''.
The observed luminosities are also not only ``kilo'' but also have some ranges.
}
A macronova was thought to be the most promising and has therefore been intensively studied.
From the theoretical side,
general relativistic simulations demonstrate the matter ejection with mass
$M_e \sim 10^{-4}$--$10^{-2} M_{\odot}$ from binary NS mergers
\citep{Hotokezaka+13,Bauswein+13,Sekiguchi+15}
(see also Refs.~\citep[][]{Kyutoku+13,Kyutoku+15} for BH-NS mergers).
The ejected matter is expected to be neutron-rich, so that
the rapid neutron capture process ($r$-process) takes place
to synthesize heavy elements such as gold, platinum, and uranium,
as a possible origin of the $r$-process nucleosynthesis \citep{Lattimer_Schramm74,Wanajo+14}.
The radioactive decay energy of the $r$-process elements heats the merger ejecta,
giving rise to a macronova \citep{Metzger+10,Kasen+13,Tanaka_Hotokezaka13}.
The $r$-process elements, in particular the lanthanides,
also increase the opacity of the ejecta to $\kappa \sim 1$--$10$ cm$^{-2}$ g$^{-1}$,
making the emission red and long-lasting
\citep{Kasen+13,Tanaka_Hotokezaka13,Tanaka+17}.

A macronova could also be powered by the central engine of an sGRB 
(see, e.g., Refs.~\citep[][]{Kisaka+15a,Kisaka+15b,Kisaka+16}).
After an NS merger, either a BH or an NS is formed. 
The central engine releases energy through a relativistic jet \citep{Kisaka_KI15}, 
disk outflows, and/or magnetar winds \citep{Fan+13,Yu+13,Metzger_Piro14},
which may be observed as prompt, extended, and plateau emissions in sGRBs
\citep{Barthelmy+05,Rowlinson+13,Gompertz+14,Kisaka+17}.
These outflows and emissions can heat the ejecta and power a macronova.

From the observational side,
a macronova candidate was detected as an infrared excess in sGRB 130603B \citep{Tanvir+13,Berger+13}.
The required mass is relatively large $> 0.02 M_{\odot}$
compared with a typical ejecta mass in the simulations
if the macronova is powered by radioactivity \citep{Hotokezaka+13b}.

\item Fourth is a radio flare \citep{Nakar_Piran11,Piran+13,Hotokezaka+16}
and the associated X-ray remnants \citep{Takami+14b}
through the interaction between the merger ejecta and the ISM.
These signals appear years later.

\end{enumerate}

Very interestingly, 
the observed EM counterparts to GW170817 do not completely follow the above expectations:
\begin{enumerate}
\item First, a faint sGRB 170817A was detected with an isotropic-equivalent energy 
$E_{\rm iso} \sim 5.35 \times 10^{46}$ erg
\citep{GW-GRB17,Fermi/GBM17,Integral17}.\footnote{
The isotropic energy $E_{\rm iso}$ is
the apparent total energy assuming that the observed emission is isotropic.}
This could arise from an off-axis sGRB jet,
but it looks like a lucky event and we should clarify
whether the signal can be produced by a typical sGRB jet or not.

\item Second, the observed optical--infrared emissions are likely a macronova,
but very bright and blue at $\sim 1$ day 
before becoming red in the following $\sim 10$ days.
Although the blue macronova could be produced by
viscously driven outflows from an accretion disk around the central engine
\citep{Fernandez_Metzger13,Kasen+15,Shibata+17},
the required ejecta mass is uncomfortably huge
$\ge 0.02 M_{\odot}$
with a small opacity $\kappa \le 0.5$ cm$^{-2}$ g$^{-1}$
to explain by $r$-process radioactivity
\citep{Tanaka+17b,Utsumi+17,Drout+17,Swift-NuSTAR17,Arcavi+17,Smartt+17,Pian+17,Kasen+17,Kasliwal+17b,Kilpatrick+17,Cowperthwaite+17,Nicholl+17,Chornock+17,McCully+17}.
The red macronova also demands a huge mass $\ge 0.03 M_{\odot}$.

These tensions motivate us to explore the contributions from jet activities to the macronova emission.
In particular, a prompt jet has to penetrate the merger ejecta 
\citep{Nagakura+14,Murguia-Berthier+14}
and inevitably injects energy into a part of the merger ejecta
to form a cocoon \citep{Nakar_Piran17,Gottlieb+17}.
We should improve analytical descriptions to calculate the observables
as functions of the jet properties because
the previous formulae are mainly for 
long GRB jets propagating in static (not expanding) stellar envelopes 
\citep{Bromberg+11,Mizuta_KI13}.

\item Third, the observed X-ray and radio afterglows are faint
with marginal detections, despite the closest sGRB ever detected.
We should check whether a typical sGRB jet is consistent with the observations or not.

\end{enumerate}

Related to all the above points,
this time, the GW observations give an important constraint on
the inclination angle $\lesssim 32^{\circ}$ ($1 \sigma$) between the binary orbital axis and the line of sight \citep{GW170817,GW170817-H0}.
Intriguingly this angle is comparable with the mean opening angle of an sGRB jet,
$\langle\Delta\theta\rangle=16^{\circ}\pm 10^{\circ}$ (1$\sigma$),
which is obtained by observing the jet break of the light curve
in addition to the non-detection of the jet break at the observation time
\citep{Fong+15}.
This finiteness of the jet size reduces the de-beaming of the off-axis emission
than the point-source case.

\begin{figure}
  \begin{center}
    \includegraphics[width=15cm]{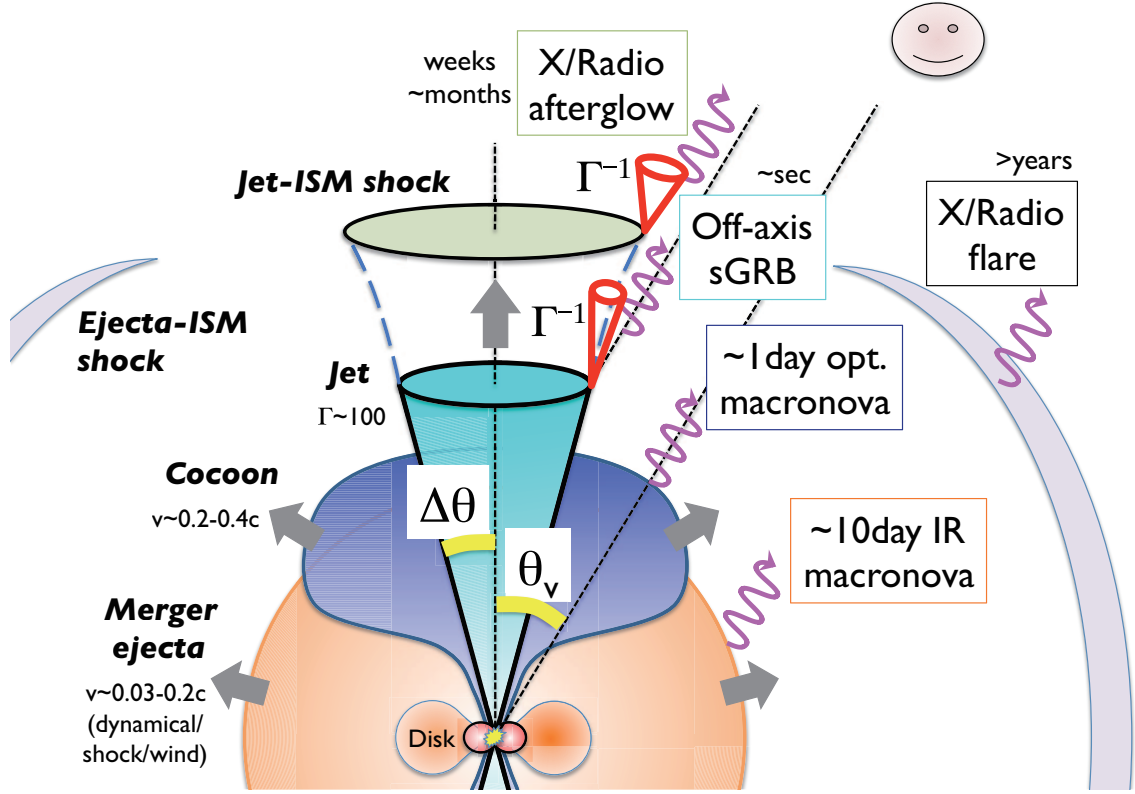}
  \end{center}
  \caption{
    Schematic figure of our unified picture.
  }
  \label{fig:model}
\end{figure}

In this paper, in order to solve the above questions,
we consider a jet associated with a neutron star merger in GW170817
and investigate its appearances in sGRB 170817A, the optical--infrared macronova, and X-ray and radio afterglows.
We then constrain the jet properties, such as the on-axis isotropic energy $E_{\rm iso}(0)$, 
opening angle $\Delta\theta$, and viewing angle $\theta_v$,
seeking whether a unified picture is possible with a typical sGRB jet or not
as in Fig.~\ref{fig:model}.

The organization of the paper is as follows.
In Sect.~\ref{sec:sGRB},
we carefully calculate the off-axis emission from a top-hat jet 
with uniform brightness and a sharp cutoff
to encompass the allowed parameter region in the plane of $E_{\rm iso}(0)$--$\Delta\theta$,
based on the formulation of \citet{KI_Nakamura01}
(see also Appendix~\ref{sec:App}).
In Sect.~\ref{sec:breakout},
we consider the jet propagation in the merger ejecta 
to derive the breakout conditions
taking the expanding motion of the merger ejecta into account.
In Sect.~\ref{sec:MN}, we calculate the expected macronova features, 
such as the flux, duration, and expansion velocity,
by improving the analytical descriptions.
In Sect.~\ref{sec:afterglow}, we estimate
the rise times and fluxes of the X-ray and radio afterglows
to constrain the jet properties and the ambient density.
In Sect.~\ref{sec:discuss}, we discuss alternative models,
and also implications for future observations of the radio flares and X-ray remnants.
Sect.~\ref{sec:sum} is devoted to the summary.
The latest observations made since submission are interpreted in Sect.~\ref{sec:new}.

\section{sGRB 170817A from an off-axis jet}\label{sec:sGRB}

The observed sGRB 170817A \citep{GW-EM17,Fermi/GBM17,Integral17} constrains
the properties of a jet associated with GW170817.
Emission from the jet is beamed into a narrow (half-)angle $\sim 1/\Gamma$
where $\Gamma$ is the Lorentz factor of the jet,
while off-axis de-beamed emission is also inevitable outside $\sim 1/\Gamma$
as a consequence of the relativistic effect
(see Fig.~\ref{fig:model}).
To begin with, we consider the most simple 
top-hat jet with uniform brightness and a sharp edge 
(see Sect.~\ref{eq:prompt} for the other cases).
For a top-hat jet,
we can easily calculate the isotropic energy $E_{\rm iso}(\theta_v)$
as a function of the viewing angle $\theta_v$
by using the formulation of \citet{KI_Nakamura01}
and Appendix~\ref{sec:App}.
Even if the observed sGRB is not the off-axis emission from a top-hat jet,
we can put the most robust upper limit on the on-axis 
isotropic energy $E_{\rm iso}(0)$ of a jet,
whatever the jet structure and the emission mechanism is.

\subsection{Isotropic energy}

The emission from a top-hat jet
is well approximated by that from a uniform thin shell 
with an opening angle $\Delta\theta$.
We can analytically obtain
the observed spectral flux in Eqs.~(\ref{eq:Fnu_App}) and (\ref{eq:deltaT}) 
\citep{KI_Nakamura01} as
\begin{eqnarray}
F_{\nu}(T)=\frac{2 r_0 c A_0}{D^2}
\frac{\Delta\phi(T) f\{\nu\Gamma[1-\beta\cos\theta(T)]\}}
{\Gamma^2[1-\beta\cos\theta(T)]^2}.
\end{eqnarray}
The isotropic energy is obtained
by numerically integrating the above equation with time and frequency as
$E_{\rm iso}(\theta_v) \propto 
\int_{T_{\rm start}}^{T_{\rm end}} dT
\int_{\nu_{\min}}^{\nu_{\max}} d\nu\,
F_{\nu}(T)$ in Eq.~(\ref{eq:Eiso_App}).
If the emission comes from multiple jets,
they usually overlap with each other, but
we can simply add all the isotropic energy\footnote{
It is not so simple to calculate 
the isotropic luminosity
because it depends on the degree of the overlap of pulses,
which depends not only on the viewing angle but also on the pulse structure
\citep{Yamazaki+02,Yamazaki+04}.
}
assuming that the jets have similar $\Delta\theta$ and $\Gamma$.

\begin{figure}
  \begin{center}
    \includegraphics[width=10cm]{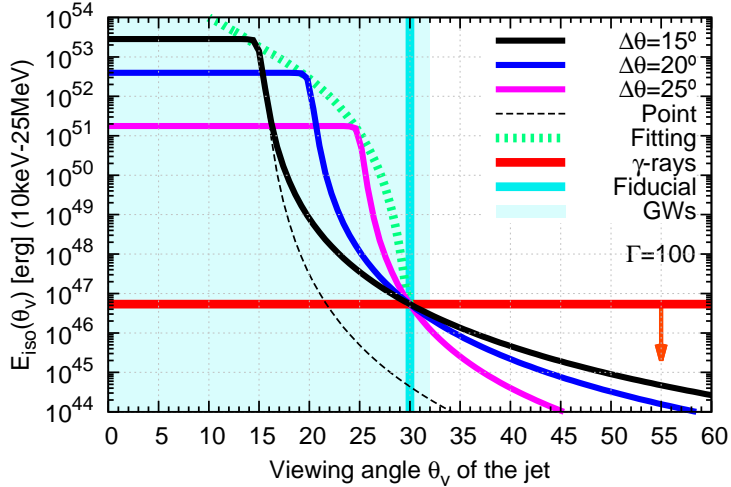}
  \end{center}
  \caption{
    Isotropic energy $E_{\rm iso}(\theta_v)$
    as a function of the viewing angle $\theta_v$
    for the opening angles of the top-hat jet 
    $\Delta \theta=15^{\circ}, 20^{\circ}, 25^{\circ}$
    with a Lorentz factor $\Gamma=100$
    calculated with the equations in the appendix.
    For a viewing angle within $\Delta \theta < \theta_v \lesssim 2\Delta\theta$,
    the isotropic energy decreases slowly as Eq.~(\ref{eq:Eiso}),
    roughly following $E_{\rm iso} \propto (\theta_v-\Delta\theta)^{-4}$,
    not $E_{\rm iso} \propto (\theta_v-\Delta\theta)^{-6}$ 
    like a point source ({\it black dashed line}).
    We normalize $E_{\rm iso}(\theta_v=30^{\circ}) = 5.35 \times 10^{46}$ erg
    ({\it red horizontal line}),
    as observed by {\it Fermi}/GBM and INTEGRAL \cite{GW-EM17,Fermi/GBM17,Integral17},
    at the fiducial viewing angle $\theta_v = 30^{\circ}$ ({\it cyan vertical line}),
    which is consistent with the inclination angle
    $\lesssim 32^{\circ}$ obtained from GWs \cite{GW170817,GW170817-H0}.
    The envelope of $E_{\rm iso}(\theta_v=\Delta\theta)$ at the jet edge
    is also plotted with the fitting formula in Eq.~(\ref{eq:upper})
    ({\it green dotted line}).
  }
  \label{fig:Eiso}
\end{figure}

In Fig.~\ref{fig:Eiso}, we calculate the isotropic energy
as a function of the viewing angle of a jet
with opening angles $\Delta\theta=15^{\circ}, 20^{\circ}, 25^{\circ}$
and $\Gamma=100$. 
We normalize $E_{\rm iso}(\theta_v=30^{\circ}) = 5.35 \times 10^{46}$ erg,
as observed by {\it Fermi}/GBM and INTEGRAL \cite{GW-EM17,Fermi/GBM17,Integral17},
at the fiducial viewing angle $\theta_v=30^{\circ}$,
which is consistent with the inclination angle
$\lesssim 32^{\circ}$ between the binary orbital axis and the line of sight
obtained from GWs \cite{GW170817,GW170817-H0}.

The important point in Fig.~\ref{fig:Eiso} is that
the viewing angle dependence of $E_{\rm iso}(\theta_v)$
for a jet with a finite opening angle $\Delta\theta > 1/\Gamma$
is quite different from the point-source case.
For a point source, there is a well-known relation 
$E_{\rm iso}(\theta_v) \propto \delta(\theta_v)^3$
between the isotropic energy $E_{\rm iso}(\theta_v)$
and the viewing angle $\theta_v$, or the Doppler factor
$\delta(\theta_v)=1/\Gamma(1-\beta\cos\theta_v)$.
However, this relation is not applicable 
if the jet size is finite and larger than $\Delta\theta > 1/\Gamma$.
As shown in Fig.~\ref{fig:Eiso} and Eqs.~(\ref{eq:Eiso1}) and (\ref{eq:Eiso2}),
the observed isotropic energy $E_{\rm iso}(\theta_v)$ is constant 
if the viewing angle is within the opening angle $\Delta \theta$.
Outside $\Delta\theta$,
the relation is initially shallower than the point-source case;
i.e., if the viewing angle is within twice the opening angle
$\Delta\theta < \theta_v \lesssim 2 \Delta\theta$,
the relation is approximately given by
\begin{eqnarray}
E_{\rm iso}(\theta_v) \propto {\tilde \delta}(\theta_v)^2
\propto 
\left[1+\Gamma^2 (\theta_v-\Delta\theta)^2\right]^{-2},
\label{eq:Eiso}
\end{eqnarray}
where the modified Doppler factor is
\begin{eqnarray}
\tilde \delta(\theta_v)=\frac{1}{\Gamma[1-\beta\cos(\theta_v-\Delta\theta)]}
\simeq \frac{2\Gamma}{1+\Gamma^2 (\theta_v-\Delta\theta)^2},
\label{eq:Doppler}
\end{eqnarray}
and we assume $\Gamma \gg 1$ and $\theta_v-\Delta\theta \ll 1$
in the last equality.
This is roughly $E_{\rm iso}(\theta_v) \propto (\theta_v-\Delta\theta)^{-4}$,
which is different from the point-source case
$E_{\rm iso}(\theta_v) \propto \delta(\theta_v)^3 \propto \theta_v^{-6}$
(see the dashed line in Fig.~\ref{fig:Eiso}).
The reason for the difference is that the flux to the observer is dominated by
the jet edge, not the jet center.
For a large enough viewing angle, i.e.,
$\theta_v \gtrsim 2\Delta \theta$,
the relation goes back to the point-source case.

Guided by the analytic equation~(\ref{eq:Eiso}),
we fit the envelope of $E_{\rm iso}(\theta_v=\Delta\theta)$ at the jet edge
in Fig.~\ref{fig:Eiso}.
This gives an upper limit on the on-axis isotropic energy
of a jet associated with sGRB 170817A
observed by {\it Fermi}/GBM and INTEGRAL \citep{GW-EM17,Fermi/GBM17,Integral17} as
\begin{eqnarray}
E_{\rm iso}(0) \le 5.35 \times 10^{46}\,{\rm erg}
\left[1+\Gamma^2(\theta_v-\Delta\theta)^2\right]^{2.3},
\label{eq:upper}
\end{eqnarray}
which is applicable for $\Gamma^{-1} \ll \Delta\theta$ and $\Delta\theta < \theta_v \lesssim 2 \Delta\theta$.

\begin{figure}
  \begin{center}
    \includegraphics[width=7.5cm]{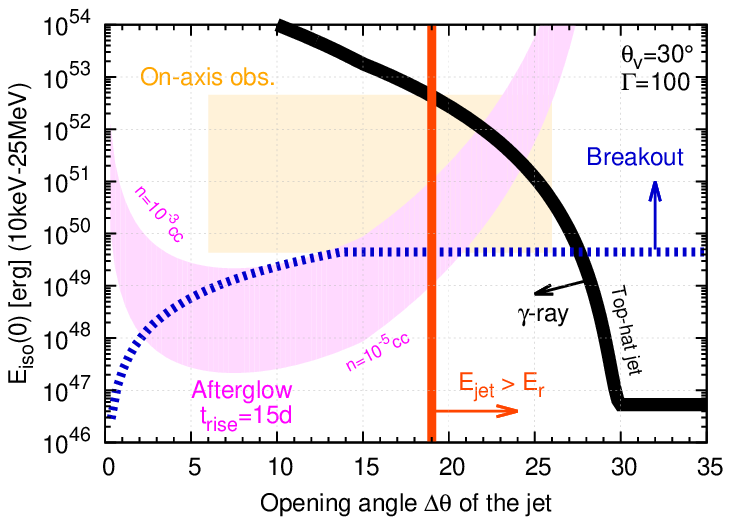}
    \includegraphics[width=7.5cm]{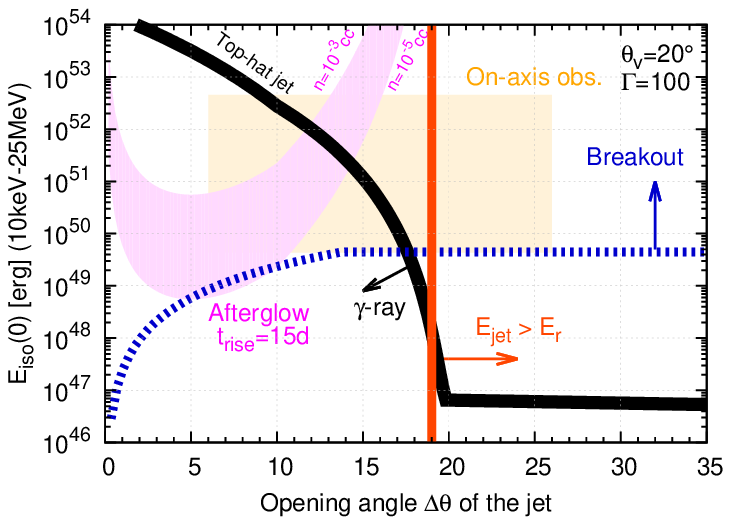}
  \end{center}
  \caption{
    The on-axis isotropic energy $E_{\rm iso}(0)$
    versus the opening angle $\Delta\theta$ of the jet
    for the fiducial viewing angles $\theta_v=30^{\circ}$ ({\it left}) 
    and $20^{\circ}$ ({\it right}).
    We plot the line for (and constraints by) a top-hat jet with $\Gamma=100$
    that explains sGRB 170817A observed by {\it Fermi}/GBM and INTEGRAL 
    \citep{GW-EM17,Fermi/GBM17,Integral17}
    ({\it black thick line}; Eq.~(\ref{eq:upper})),
    the jet breakout condition ({\it blue dotted line}; Sect.~\ref{sec:breakout}),
    the condition for the jet energy to dominate the radioactive energy 
    for the blue macronova
    ({\it red vertical line}; Sect.~\ref{sec:MN}),
    the region for the rise time $t_{\rm rise}=15$ d of the X-ray/radio afterglows
    with the ambient density $n=10^{-5}$--$10^{-3}$ cm$^{-3}$
    ({\it magenta curved region}; Sect.~\ref{sec:afterglow}),
    and the observed region for $E_{\rm iso}(0)$ and $\Delta\theta$ of the past sGRBs
    that are thought to be on-axis ({\it orange square}; Sect.~\ref{sec:sGRB}).
  }
  \label{fig:Eiso-dth}
\end{figure}

\begin{figure}
  \begin{center}
    \includegraphics[width=7.5cm]{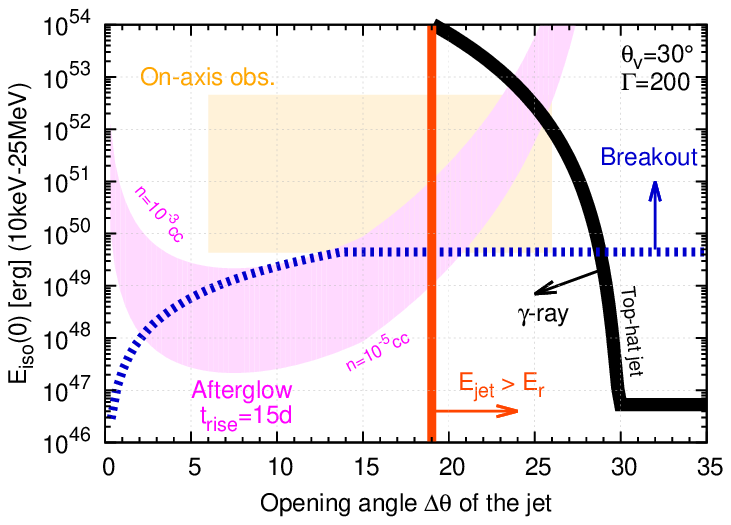}
    \includegraphics[width=7.5cm]{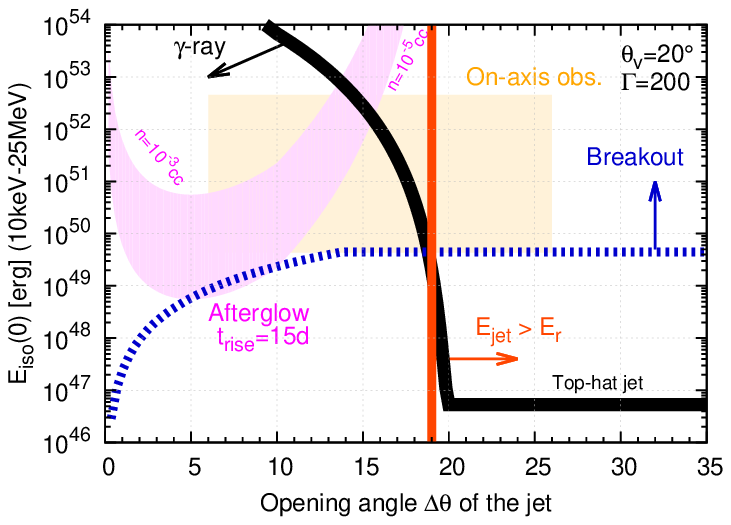}
  \end{center}
  \caption{
    Same as Fig.~\ref{fig:Eiso-dth} except for $\Gamma=200$.
  }
  \label{fig:Eiso-dth2}
\end{figure}

In Fig.~\ref{fig:Eiso-dth} (for $\Gamma=100$; {\it black thick line})
and Fig.~\ref{fig:Eiso-dth2} (for $\Gamma=200$; {\it black thick line}),
we plot the upper limit on the on-axis isotropic energy in Eq.~(\ref{eq:upper})
as a function of the opening angle $\Delta\theta$
with the fiducial viewing angles $\theta_v=30^{\circ}$ and $20^{\circ}$,
which are consistent with
the inclination angle $\lesssim 32^{\circ}$ 
obtained from GWs \citep{GW170817,GW170817-H0}.
We adopt two cases $\Gamma=100$ and $200$ since
the Lorentz factor of sGRBs is not well constrained.
Although much larger lower limits $\Gamma \gtrsim 1000$
have been derived for sGRB 090510 detected by the {\it Fermi}/LAT
\citep{Ackermann+10},
these limits rely on the one-zone model,
and are reduced by a factor of several
in multi-zone models \citep{Aoi+10,Zou+11,Hascoet+12}.
As for long GRBs,
\citet{Hascot+14} obtain density-dependent lower limits $\Gamma > 40$--$300$,
and \citet{Nava+17} obtain upper limits
$\Gamma < 200$ for a homogeneous density medium 
and $\Gamma < 100$--$400$ for a wind-like medium.

In Figs.~\ref{fig:Eiso-dth} and \ref{fig:Eiso-dth2}
({\it the vertical range of the orange square}),
we also plot the range of the isotropic energies,
$E_{\rm iso}= 4.33 \times 10^{49}$--$4.54 \times 10^{52}$ erg,
for the past sGRBs
that are thought to be on-axis
because they satisfy 
the $E_{p}$--$E_{\rm iso}$ (Amati)
and $E_{p}$--$L_{\rm iso}$ (Yonetoku)
relations \cite{Tsutsui:2013,Amati02,Yonetoku04}.
As we can see from Figs.~\ref{fig:Eiso-dth} and \ref{fig:Eiso-dth2},
a top-hat jet with typical on-axis isotropic energy $E_{\rm iso}(0)$ can explain 
the faint sGRB 170817A
if the viewing angle of the jet edge is in the range
\begin{eqnarray}
3^{\circ} \left(\frac{\Gamma}{100}\right)^{-1} 
< \theta_v -\Delta\theta 
< 11^{\circ} \left(\frac{\Gamma}{100}\right)^{-1},
\label{eq:allow}
\end{eqnarray}
with Eq.~(\ref{eq:upper}).

In Figs.~\ref{fig:Eiso-dth} and \ref{fig:Eiso-dth2}
({\it the horizontal range of the orange square}),
we also plot the range of the mean opening angle 
$\langle\Delta\theta\rangle=16^{\circ}\pm 10^{\circ}$ (1$\sigma$),
which is obtained by observing the jet break of the light curve
in addition to the non-detection of the jet break at the observation time
\citep{Fong+15}.
A top-hat jet for sGRB 170817A
can also take these typical opening angles 
unless $\theta_v > 26^{\circ} + 11^{\circ}(\Gamma/100)^{-1}$
or $\theta_v < 6^{\circ} + 3^{\circ}(\Gamma/100)^{-1}$.

\subsection{Spectrum}\label{sec:spec}

Although spectral information is important for discriminating models,
sGRB 170817A is faint and 
it is difficult to draw a robust conclusion based on its spectrum.
Detailed analysis of sGRB 170817A revealed two components to the burst:
a main pulse with $\sim 0.6$ s and a weak tail
with $34\%$ the fluence of the main pulse \citep{GW-EM17}.
The main pulse is best fitted with a Comptonized spectrum
with a power-law photon index of $-0.62 \pm 0.40$
and peak energy $E_{p}=185 \pm 62$ keV,
while the weak tail has a softer blackbody spectrum with $k_B T=10.3 \pm 1.5$ keV
\citep{GW-EM17}.
The de-beamed emission from an off-axis top-hat jet
tends to have a low spectral peak at
\begin{eqnarray}
E_{p}(\theta_v) \sim 
\left[\frac{\tilde \delta (\theta_v)}{\tilde \delta(0)}\right] E_{p}(0)
\sim 10\,{\rm keV}\, 
\left[\frac{\Gamma(\theta_v-\Delta\theta)}{10}\right]^{-2}
\left[\frac{E_{p}(0)}{\rm MeV}\right],
\label{eq:Ep}
\end{eqnarray}
where the Doppler factor 
$\tilde \delta (\theta_v)$ is given by Eq.~(\ref{eq:Doppler}).
This is consistent with the observed $E_{p}$ of the main pulse within $3\sigma$
and also with the $k_B T$ of the weak tail.

On the other hand,
if we believe that the central value of the peak energy
$E_{p}=185 \pm 62$ keV is correct for the main pulse,
the on-axis peak energy lies outside
the $E_{p}$-$E_{\rm iso}$ (Amati) and
$E_{p}$-$L_{\rm iso}$ (Yonetoku) relations
\citep{Tsutsui:2013,Amati02,Yonetoku04},
implying a different emission mechanism.
In any case, we should keep in mind that
GRB 170817A could have been $\sim 30\%$ dimmer
before falling below the on-board triggering threshold \citep{GW-GRB17}.
It is also detected just before entering the South Atlantic Anomaly.
In addition, the well-known correlation between $E_p$ and the peak luminosity for each pulse
possibly biases the peak energy toward a high value for a tip-of-the-iceberg event.

The spectral shape above the peak energy
is also not measured well, so that
the compactness problem does not give a strong limit on
the Lorentz factor.
The most conservative case is that
the spectrum is sharply cut off above the peak energy.
In this case, the electron and positron pairs are not created
if the peak energy in the comoving frame $\sim \Gamma E_p(0)$
is less than the electron rest mass energy $m_e c^2$,
which only gives a weak constraint on the Lorentz factor:\footnote{
  The opacity due to electrons associated with protons
  typically gives a lower limit of about $\Gamma \gtrsim 50$.
  }
\begin{equation}
  \Gamma \gtrsim \frac{E_p(0)}{m_e c^2}
  \sim 1 \left[\frac{E_p(0)}{\rm MeV}\right].
\end{equation}
With Eq.~(\ref{eq:Ep}),
this gives an upper limit on the viewing angle,
$\theta_v-\Delta \theta < 10
\left[E_p(\theta_v)/10\,{\rm keV}\right]^{-1}
\left[\Gamma(\theta_v-\Delta\theta)/10\right]^{-1}$.
On the other hand,
if we assume that
the spectrum above the peak energy is exponentially cut off,
the optical depth exceeds unity
unless the minimum Lorentz factor is
$\Gamma \gtrsim 100 \left[\Gamma(\theta_v-\Delta\theta)/10\right]^{4/3}$
\cite{Kasliwal+17}.
The cutoff shape, e.g., the index $\lambda$
in the cutoff $\exp[-(E/E_p)^{\lambda}]$,
depends on the emission mechanism,
which is still unknown and future observations are anticipated.
Note also that the optical depth is
angle dependent near the photosphere \citep{Abramowicz+91},
and the Doppler factor is not the only control parameter.


\section{Jet breakout}\label{sec:breakout}

An NS--NS merger gives rise to matter ejection
with masses $M_e \sim 10^{-4}$--$10^{-2}M_{\odot}$
and velocities $0.1$--$0.3c$
in a quasispherical manner before the jet launch
\citep{Hotokezaka+13,Bauswein+13,Sekiguchi+15}.
Simulations of numerical relativity actually show that
the mass of $M_e \sim 10^{-2} M_{\odot}$ is ejected \citep{Shibata+17b}
from a system similar to that observed by GWs
with the NS masses $1.17$--$1.60 M_{\odot}$
and total mass $2.74^{+0.04}_{-0.01} M_{\odot}$ \citep{GW170817}.
The GWs place a 90\% upper limit on the tidal deformability
$\Lambda_1 \lesssim 1500$ and $\Lambda_2 \lesssim 3000$
in the low-spin case (see Fig.~5 in Ref.~\citep[][]{GW170817}),
disfavoring an equation of state (EOS) for less-compact NSs such as MS1.
The compact, deep gravitational potential strengthens 
the shock heating, rather than the tidal torque,
at the onset of the merger,
enhancing mass ejection to the orbital axis.
In addition to the dynamical mass ejection,
neutrino-driven winds \citep{Dessart+09,Perego+14,Fujibayashi+17}
and more importantly 
viscously driven outflows from an accretion disk
eject mass to the jet axis 
\citep{Fernandez_Metzger13,Fernandez+15,Just+15,Kasen+15,Siegel_Metzger17,Lippuner+17,Shibata+17},
which is increased by mass asymmetry,
although a robust conclusion should wait for
general relativistic simulations with magnetic fields
\citep{Kiuchi+14,Kiuchi+15,Kiuchi+17}.

The jet has to penetrate the merger ejecta to be observed as the sGRB
\citep{Nagakura+14,Murguia-Berthier+14,Gottlieb+17}.
In particular, the breakout time $t_{\rm br}$ 
should be less than the delay time $\Delta T_{0} \sim 2$ s
of the sGRB 170817A from the GW detection.
Note that the delay time is the sum 
\begin{eqnarray}
\Delta T_0=t_{j}+t_{\rm br}+(T_{\rm start}-T_0),
\end{eqnarray}
of the launch time of the jet $t_j$,
the breakout time $t_{\rm br}$,
and the starting time of a single pulse in Eq.~(\ref{eq:Tstart}),
\begin{eqnarray}
T_{\rm start}-T_0 \sim 
\frac{r_0}{c\beta} \left[1-\beta\cos(\max[0,\theta_v-\Delta\theta])\right]
\sim 2\,{\rm s}\,
\left(\frac{r_0}{10^{13}\,{\rm cm}}\right)
\left(\frac{\theta_v-\Delta\theta}{0.1}\right)^2,
\label{eq:TstartT0}
\end{eqnarray}
due to the difference between the straight path to the center and
the path via the emission site at a radius $r_0$,
i.e., a kind of curvature effect,
where we assume $\theta_v-\Delta\theta > 1/\Gamma$ in the last equality.
Even if the breakout is very fast,
the sGRB may not start immediately.
The breakout condition $t_{\rm br}<\Delta T_{0} \sim 2$ s
is the necessary condition for the sGRB.
Hereafter we assume $t_j \ll \Delta T_{0}$ for simplicity 
(see Sect.~\ref{sec:tj} for discussions).

The jet head velocity is determined by the ram pressure 
balance between the shocked jet and the shocked ejecta, 
both of which are given by the pre-shock quantities
through the shock jump conditions,
\begin{eqnarray}
h_j \rho_j c^2 \Gamma_{jh}\beta_{jh}^2+P_j =h_e \rho_e c^2\Gamma_{he}^2\beta_{he}^2+P_e,
\end{eqnarray}
where $\Gamma_{AB}=\Gamma_A\Gamma_B(1-\beta_A\beta_B)$
is the relative Lorentz factor between the jet head ($h$) and
the jet ($j$) or the ejecta ($e$)
and $\beta_{AB}=(\beta_A-\beta_B)/(1-\beta_A\beta_B)$
is the corresponding relative velocity
(see, e.g., Refs.~\citep[][]{Marti+97,Matzner03}).
We can neglect the internal pressure of the jet $P_j$ and the ejecta $P_e$.
Then the relative velocity between the jet head and the ejecta is 
\begin{eqnarray}
\beta_h-\beta_e=\frac{\beta_j-\beta_e}{1+{\tilde L}^{-1/2}},
\label{eq:jheadv}
\end{eqnarray}
where the ratio of the energy density between the jet and the ejecta is
\begin{eqnarray}
\tilde L \equiv \frac{h_j \rho_j \Gamma_j^2}{h_e \rho_e \Gamma_e^2}
\simeq \frac{L_j}{\Sigma_j \rho_e c^3}.
\label{eq:tildeL}
\end{eqnarray}
In the last equality, we assume the cold ejecta $h_e=1$,
and use the jet cross section $\Sigma_j=\pi r_j^2$ and the (one-sided) jet luminosity $L_j$.
The jet luminosity is given by the on-axis isotropic energy, 
opening angle, and duration of the jet activity $t_{\rm dur}$ as
\begin{eqnarray}
L_j \sim \frac{\Delta\theta^2}{4} \frac{E_{\rm iso}(0)}{\epsilon_{\gamma} t_{\rm dur}}
\sim 3 \times 10^{50}\,{\rm erg}\,{\rm s}^{-1}
\left(\frac{\Delta\theta}{0.3}\right)^2
\left(\frac{E_{\rm iso}(0)/\epsilon_{\gamma}}{3\times 10^{52}\,{\rm erg}}\right)
\left(\frac{t_{\rm dur}}{2\,{\rm s}}\right)^{-1},
\label{eq:Lj}
\end{eqnarray}
where $\epsilon_{\gamma} \sim 0.1$ is the $\gamma$-ray efficiency.

The ejecta density at time $t$ is
\begin{eqnarray}
\rho_e \sim \frac{3M_e}{4\pi (c\beta_e t)^3}
\sim 3\,{\rm g}\,{\rm cm}^{-3}
\left(\frac{M_e}{0.01M_{\odot}}\right)
\left(\frac{\beta_e}{0.2}\right)^{-3}
\left(\frac{t}{2\,{\rm s}}\right)^{-3}.
\label{eq:rhoe}
\end{eqnarray}
The dynamical mass ejection to the orbital axis 
is primarily caused by the shock heating at the onset of the merger
rather than the tidal torque.
While the ejected mass to the orbital axis is relatively smaller than
that to the orbital plane \citep{Sekiguchi+15},
the GWs disfavor an EOS for less-compact NSs \citep{GW170817},
implying efficient shock heating \citep{Hotokezaka+13,Sekiguchi+15}.
Viscous outflows from an accretion disk also add mass to the jet axis
\citep{Fernandez_Metzger13,Fernandez+15,Just+15,Kasen+15,Siegel_Metzger17,Lippuner+17,Shibata+17}.
Most of the dynamical ejecta has a velocity of $\beta_e \sim 0.2$,
although the head of the dynamical ejecta is rapid
\citep{Just+15} even up to ultrarelativistic speeds \citep{Kyutoku+14}.
The velocity of viscous outflows is thought to be moderate 
$\beta_e \sim 0.03$--$0.1$.

First we consider the case that the jet is not collimated.
Then the cross section of the jet at the breakout time is
$\Sigma_j \sim \pi (\Delta\theta c\beta_{e} t_{\rm br})^2$,
so that Eqs.~(\ref{eq:tildeL}), (\ref{eq:Lj}), and (\ref{eq:rhoe}) yield
\begin{eqnarray}
\tilde L \sim 0.1
\left(\frac{E_{\rm iso}(0)/\epsilon_{\gamma}}{3\times 10^{52}\,{\rm erg}}\right)
\left(\frac{t_{\rm dur}}{2\,{\rm s}}\right)^{-1}
\left(\frac{t_{\rm br}}{2\,{\rm s}}\right)
\left(\frac{\beta_e}{0.2}\right)
\left(\frac{M_e}{0.01M_{\odot}}\right)^{-1},
\end{eqnarray}
where $t_{\rm dur}$ is the jet duration.
The breakout time is determined by 
the condition that the jet head moves through the ejecta size,
\begin{eqnarray}
c \beta_e t_{\rm br}\sim c (\beta_h-\beta_e)t_{\rm br},
\label{eq:breakout}
\end{eqnarray}
because the jet head velocity is slow in the early phase when the ejecta density is high.
This yields $\tilde L \simeq \beta_e^2/(1-2\beta_e)^2$ with Eq.~(\ref{eq:jheadv})
for $\beta_j \simeq 1$,
and therefore
\begin{eqnarray}
t_{\rm br} \sim 2\,{\rm s}
\left(\frac{E_{\rm iso}(0)/\epsilon_{\gamma}}{3\times 10^{52}\,{\rm erg}}\right)^{-1}
\left(\frac{t_{\rm dur}}{2\,{\rm s}}\right)
\left(\frac{\beta_e/(1-2\beta_e)^2}{0.2/0.6^2}\right)
\left(\frac{M_e}{0.01M_{\odot}}\right).
\label{eq:tbr1}
\end{eqnarray}
The parameter dependence is different from 
that for the jet breakout from a stellar envelope
\citep{Nakar_Piran17,Lazzati+17a},
because the merger ejecta is moving outward and 
the jet head velocity at the breakout automatically becomes
comparable to the ejecta velocity $\beta_h \sim 2\beta_e$,
not very fast or slow as in the case of the stellar breakout.

Next we consider the collimated case.
The shocked jet and the shocked ejecta go sideways from the jet head
and form a cocoon \citep{Ramirez-Ruiz+02,Bromberg+11,Mizuta_KI13}.
If the cocoon pressure becomes higher than the jet pressure,
the jet is collimated and the propagation is modified.
The collimated jet dynamics was studied in the context of long GRBs
\citep{Bromberg+11,Mizuta_KI13,Harrison+17}.
The numerically calibrated equation for the jet head position is obtained
in \citet{Mizuta_KI13} as
\begin{eqnarray}
z_h \sim 1.4 \times 10^{10}\,{\rm cm}
\left(\frac{t}{1\,{\rm s}}\right)^{3/5}
\left(\frac{L_j}{10^{51}\,{\rm erg}\,{\rm s}^{-1}}\right)^{1/5}
\left(\frac{\rho_e}{10^{3}\,{\rm g}\,{\rm cm}^{-3}}\right)^{-1/5}
\left(\frac{\Delta\theta_0}{0.1}\right)^{-4/5}.
\end{eqnarray}
Substituting Eqs.~(\ref{eq:Lj}) and (\ref{eq:rhoe}) and 
$z_h \sim c\beta_e t_{\rm br}$ into the above equation,
we obtain the breakout time of the collimated jet as
\begin{eqnarray}
t_{\rm br2} \sim 3\,{\rm s}
\left(\frac{\Delta\theta}{0.3}\right)^{2}
\left(\frac{E_{\rm iso}(0)/\epsilon_{\gamma}}{3\times 10^{52}\,{\rm erg}}\right)^{-1}
\left(\frac{t_{\rm dur}}{2\,{\rm s}}\right)
\left(\frac{\beta_e}{0.2}\right)^2
\left(\frac{M_e}{0.01M_{\odot}}\right)
\left(\frac{\Delta\theta_0/\Delta\theta}{3}\right)^4,
\label{eq:tbr2}
\end{eqnarray}
where we take into account that
the opening angle after the breakout becomes narrower than the initial one
$\Delta \theta \sim \Delta \theta_0/3$
because of the acceleration at the jet breakout \citep{Mizuta_KI13}.\footnote{
\citet{Mizuta_KI13} shows the ratio
$\Delta_0/\Delta\theta \sim 5$
in the case of the jet breakout from the stellar envelope.
Since the merger ejecta has a different density profile from the stellar envelope,
the ratio could be different.
Here we take a small ratio for conservative estimates.}
The condition for the collimation is given by $\tilde L \le \Delta \theta_0^{-4/3}$ \citep{Bromberg+11}.
The breakout time is given by the shorter of 
Eqs.~(\ref{eq:tbr1}) and (\ref{eq:tbr2}).

In Fig.~\ref{fig:Eiso-dth} ({\it blue dotted line}), we plot the condition
for the breakout to occur before the sGRB $t_{\rm br}<\Delta T_0 \sim 2$ s.
We can see that the breakout is possible for typical sGRBs.
In applying Eqs.~(\ref{eq:tbr1}) and (\ref{eq:tbr2}),
we should be careful about the duration of the jet activity $t_{\rm dur}$.
In the on-axis case,
the jet duration $t_{\rm dur}$ is usually equal to the observed sGRB duration $T_{90}$.
This is not always the case for the off-axis jet.
The apparent duration measured by observers is given by
\begin{eqnarray}
T_{90} \sim \max[t_{\rm dur}, \Delta T],
\end{eqnarray}
where $\Delta T$ is the duration of a single pulse in Eq.~(\ref{eq:scale}),
and if $\Delta\theta<\theta_v\lesssim 2\Delta\theta$ and $1/\Gamma \ll \Delta\theta$,
\begin{eqnarray}
\Delta T \sim \frac{r_0}{c\beta}
\left[1-\beta\cos(\theta_v-\Delta\theta)\right]
\sim 2\,{\rm s}\,
\left(\frac{r_0}{10^{13}\,{\rm cm}}\right)
\left(\frac{\theta_v-\Delta\theta}{0.1}\right)^2.
\label{eq:duration}
\end{eqnarray}
Even if the jet duration is much shorter than $t_{\rm dur} \ll 2$ s,
the observed duration may be $T_{90} \sim 2$ s as observed.\footnote{
Note that the duration in Eq.~(\ref{eq:duration})
is comparable to the starting time of a pulse in Eq.~(\ref{eq:TstartT0}).
If this is the reason for the similarity of $T_{90}$ and $\Delta T_0$ in sGRB 170817,
the breakout time $t_{\rm br}$ should be shorter than 
$\Delta T_0 \sim 2$ s.
The similarity is also realized if $t_{\rm br} \sim t_{\rm dur} \sim 2$ s.}
To be conservative, we take $t_{\rm dur} > 0.03$ s, 
which is nearly the shortest duration of the observed sGRBs.

\section{Blue macronova powered by a jet?}\label{sec:MN}

The jet propagating through the merger ejecta
injects energy into the cocoon, 
which is the mixed sum of the shocked jet and the shocked ejecta.
The injected energy accelerates a part of the merger ejecta,
and also heats the ejecta, contributing to the macronova emission
\citep{Kisaka+15a,Kisaka_KI15,Kisaka+17,Nakar_Piran17,Gottlieb+17}.
We consider the uncollimated case, which is mainly relevant to our case.
The injected energy from two-sided jets is estimated
from Eqs.~(\ref{eq:Lj}) and (\ref{eq:tbr1}) as
\begin{eqnarray}
E_{\rm inj} \sim 2 L_j t_{\rm br}
\sim 1 \times 10^{51}\,{\rm erg}
\left(\frac{\Delta\theta}{0.3}\right)^2
\left(\frac{\beta_e/(1-2\beta_e)^2}{0.2/0.6^2}\right)
\left(\frac{M_e}{0.01M_{\odot}}\right),
\label{eq:Einj}
\end{eqnarray}
which is interestingly independent of the jet luminosity.
The shocked fraction of the merger ejecta is 
$f_c \sim (\beta_{\perp}/\beta_h)^2/2 \sim (\beta_{\perp}/2\beta_e)^2/2$,
where the lateral velocity of the shock is
$\beta_{\perp} \sim \sqrt{{E_{\rm inj}}/{f_c M_ec^2}}$, and therefore
\begin{eqnarray}
\beta_{\perp} \sim \left(8\beta_e^2 \frac{E_{\rm inj}}{M_ec^2}\right)^{1/4}
\sim 0.4 
\left(\frac{\Delta\theta}{0.3}\right)^{1/2}
\left(\frac{\beta_e^3/(1-2\beta_e)^2}{0.2^3/0.6^2}\right)^{1/4},
\label{eq:betaperp}
\end{eqnarray}
which is also interestingly independent of the ejecta mass.
This gives the cocoon velocity and mass:
\begin{eqnarray}
\beta_c &\sim& \sqrt{{\beta}_{\perp}^2+\beta_e^2},
\label{eq:betac}
\\
M_c &=& f_c M_e \sim 0.5 M_e 
\left(\frac{\Delta\theta}{0.3}\right)
\left(\frac{\beta_e (1-2\beta_e)^{2}}{0.2\cdot 0.6^2}\right)^{-1/2}.
\label{eq:Mc}
\end{eqnarray}
Note that the cocoon mass is comparable to the ejecta mass
and proportional to $M_c \propto \Delta\theta$,
not so small $\sim (\Delta\theta)^2/2 \sim 0.05 (\Delta\theta/0.3)^2$
as in the case of the jet breakout from the stellar envelope.
This is because the jet head velocity is naturally tuned to 
the ejecta velocity $\beta_h \sim 2\beta_e$ at the breakout in Eq.~(\ref{eq:breakout})
and the lateral velocity of the shock
is also comparable to the ejecta velocity $\beta_{\perp} \sim 2\beta_e$
for typical opening angles $\Delta\theta$ in Eq.~(\ref{eq:betaperp}).
The large cocoon mass with $M_c \propto \Delta\theta$ 
in the jet breakout from merger ejecta
has not been analytically pointed out so far, as far as we know.

The energy injected by the jet is released at the 
photospheric radius $r_{\rm ph} \sim c \beta_c t_{\rm MN}$ of the macronova emission
at the peak time $t_{\rm MN}$.
This is larger than the radius of the energy injection $r_{\rm br} \sim c \beta_e t_{\rm br}$,
so that the adiabatic cooling reduces the released energy as
\begin{eqnarray}
E_{\rm jet} \sim \frac{r_{\rm br}}{r_{\rm ph}} E_{\rm inj}
\sim 1.4 \times 10^{46}\,{\rm erg}
\left(\frac{t_{\rm MN}}{1\,{\rm day}}\right)^{-1}
\left(\frac{t_{\rm br}}{2\,{\rm s}}\right)
\left(\frac{\Delta\theta}{0.3}\right)^2
\left(\frac{\beta_e/(1-2\beta_e)^2}{0.2/0.6^2}\right)
\left(\frac{M_e}{0.01M_{\odot}}\right),
\label{eq:E_MN}
\end{eqnarray}
where we omit the parameter dependence of $\beta_c/\beta_e$ in the last expression.

Let us first compare the jet energy in Eq.~(\ref{eq:E_MN})
with the energy released by radioactive decays in the macronova emission.
The merger ejecta is likely neutron rich
and a possible site of $r$-process nucleosynthesis
\citep{Lattimer_Schramm74,Metzger+10}.
Synthesized nuclei are unstable
and the radioactive energy can also power a macronova
\citep{Li_Paczynski98,Kulkarni05}.
The dominant contribution to the macronova emission
is determined by the radioactive heating rate ${\dot \varepsilon}_r$
at the peak time $t_{\rm MN}$
because the energy injected before $t_{\rm MN}$ is adiabatically cooled down.
Then the radioactive energy in the macronova emission is estimated as
\begin{eqnarray}
E_{r} \sim \epsilon_{\rm th} {\dot \varepsilon}_r t_{\rm MN} M_e
\sim 1.7 \times 10^{46}\,{\rm erg}
\left(\frac{t_{\rm MN}}{1\,{\rm day}}\right)^{-0.3}
\left(\frac{M_e}{0.01M_{\odot}}\right),
\label{eq:E_r}
\end{eqnarray}
where we adopt the heating rate
${\dot \varepsilon}_r = 2\times 10^{10} (t/1\,{\rm day})^{-1.3}$ erg s$^{-1}$ g$^{-1}$,
which gives a reasonable agreement 
with nucleosynthesis calculations for a wide range of 
the electron fraction $Y_e$ \citep{Wanajo+14},
and the thermalization factor $\epsilon_{\rm th} \sim 0.5$, 
which is a (time-dependent) fraction of the decay energy deposited to the ejecta 
at $t_{\rm MN} \sim 1$ day \citep{Barnes+16,Hotokezaka+16b}.
Note that there still remain uncertainties in the released energy
$E_r$ by a factor of $2$--$3$ due to the nuclear models, in particular 
the abundance of $\alpha$-decaying trans-lead nuclei \citep{Rosswog+17}.
The jet energy in Eq.~(\ref{eq:E_MN}) 
dominates the radioactive energy, $E_{\rm jet}>E_r$,
if the opening angle is wide enough,
\begin{eqnarray}
\Delta\theta \gtrsim 19^{\circ}
\left(\frac{t_{\rm MN}}{1\,{\rm day}}\right)^{0.35}
\left(\frac{t_{\rm br}}{2\,{\rm s}}\right)^{-1/2}
\left(\frac{\beta_e/(1-2\beta_e)^2}{0.2/0.6^2}\right)^{-1/2}.
\label{eq:Ej>Er}
\end{eqnarray}
This is shown in Figs.~\ref{fig:Eiso-dth} and \ref{fig:Eiso-dth2}
({\it red vertical line})
for the breakout time $t_{\rm br} = 2$ s, 
the peak time of the macronova $t_{\rm MN}=1$ day,
and the ejecta velocity $\beta_e=0.2$.
We can see that if the viewing angle is $20 \lesssim \theta_v \lesssim 30^{\circ}$,
there is a parameter space for the jet to dominate the macronova energy,
while if $\theta_v \lesssim 20^{\circ}$,
the prompt jet alone cannot dominate the macronova energy.
Note that the ejecta mass $M_e$ is canceled in Eq.~(\ref{eq:Ej>Er}).

Now let us consider the observed macronova.
The observed temperature $T_{\rm MN}$ and luminosity $L_{\rm MN}$
suggest that the emission region is different 
between at $t_{\rm MN} \sim 1$ day and $10$ day.
In particular, the macronova is blue at $t_{\rm MN} \sim 1$ day and becomes red later
\citep{GW-EM17}.
The photospheric velocity
\begin{eqnarray}
\beta_{\rm ph} \sim \frac{1}{ct_{\rm MN}}
\sqrt{\frac{L_{\rm MN}}{4\pi \Omega \sigma T_{\rm MN}^4}}
\sim 0.3 \left(\frac{T_{\rm MN}}{8000\,{\rm K}}\right)^{-2}
\left(\frac{L_{\rm MN}}{10^{42}\,{\rm erg}\,{\rm s}^{-1}}\right)^{1/2}
\left(\frac{t_{\rm MN}}{1\,{\rm day}}\right)^{-1},
\end{eqnarray}
is $\beta_{\rm ph} \sim 0.3$--$0.4$ at $t_{\rm MN} \sim 1$ day
($T_{\rm MN} \sim 7000$--$10^4$ K, 
$L_{\rm MN} \sim 7 \times 10^{41}$--$1 \times 10^{42}$ erg s$^{-1}$)
and $\beta_{\rm ph} \sim 0.1$ at $t_{\rm MN} \sim 10$ day 
($T_{\rm MN} \sim 2000$ K, $L_{\rm MN} \sim 8 \times 10^{40}$ erg s$^{-1}$)
\citep{GW-EM17,Tanaka+17b,Arcavi+17,Smartt+17,Shappee+17,Pian+17,Kasen+17},
where $\Omega \sim 0.5$ is the fraction of the solid angle of the emission region.
The different emission regions indicate
some structures in the polar or radial direction
\citep{Smartt+17,Waxman+17}.
Such structures of the density and composition
could be shaped by the jet activities.

The blue macronova emission at $t_{\rm MN} \sim 1$ day
naturally come from the cocoon
accelerated by the jet.
This is because the photospheric velocity $\sim 0.3$--$0.4 c$ is faster than
the typical velocity of the dynamical ejecta $\sim 0.2 c$\footnote{
The fast photospheric velocity $\sim 0.3$--$0.4 c$ may still be
explained by a velocity structure of the dynamical ejecta.
}
and the disk outflows $\sim 0.03$--$0.1 c$ obtained in the numerical simulations
\citep{Fernandez_Metzger13,Fernandez+15,Just+15,Kasen+15,Siegel_Metzger17,Lippuner+17,Shibata+17},
but is consistent with the cocoon velocity
$\beta_c \sim \sqrt{{\beta}_{\perp}^2+\beta_e^2} \sim 0.4$
in Eqs.~(\ref{eq:betaperp}) and (\ref{eq:betac}).
The duration of $t_{\rm MN} \sim 1$ day
is also consistent with the diffusion time of photons in the cocoon,
\begin{eqnarray}
t_{\rm diff} \sim 
\sqrt{\frac{2\kappa M_c}{B \Omega c^2 \beta_c}}
\sim 1\,{\rm day} 
\left(\frac{\kappa}{1\,{\rm cm}^2\,{\rm g}^{-1}}\right)^{1/2} 
\left(\frac{M_c}{0.005 M_{\odot}}\right)^{1/2} 
\left(\frac{\beta_c}{0.4}\right)^{-1/2},
\label{eq:tdiff}
\end{eqnarray}
where $B \approx 13.7$ is an integration constant following \citet{Arnett82}
and $\kappa$ is the opacity.
Here the opacity is increased by the $r$-process nucleosynthesis
\citep{Kasen+13,Barnes_Kasen13,Tanaka_Hotokezaka13}
and in particular is very sensitive to the amount of lanthanide elements 
\citep{Kasen+13,Tanaka+17}.
The merger ejecta along the jet, i.e.,
the shock-heated dynamical ejecta and the disk outflows,
tend to have a large electron fraction $Y_e \sim 0.25$--$0.4$
\citep{Fernandez_Metzger13,Kasen+15,Fernandez+15,Just+15,Siegel_Metzger17,Lippuner+17,Shibata+17},
producing only $r$-process elements below the second peak.
This leads to a small opacity $\kappa \sim 0.1$--$1$ cm$^{2}$ g$^{-1}$
\citep{Metzger_Fernandez14,Kasen+15},
compared with that of the dynamical ejecta $\kappa \sim 10$ cm$^{2}$ g$^{-1}$.
An intermediate opacity $\kappa \sim 1$ cm$^{2}$ g$^{-1}$ could also be realized
by the turbulent mixing of the dynamical ejecta and the disk outflows in the cocoon.

The radiated energy of the blue macronova at $t_{\rm MN}\sim 1$ day 
is too large $\sim 7 \times 10^{46}$ erg to be explained
by the radioactivity if the ejecta mass is typical $M_e \sim 0.01M_{\odot}$
as in the numerical simulations \citep{Hotokezaka+13,Sekiguchi+15}.
The radioactive model requires large ejecta mass $M_e \sim 0.02 M_{\odot}$
(for large energy)
as well as a small opacity $\kappa \sim 0.1$ cm$^{2}$ g$^{-1}$
(for a $t_{\rm MN}\sim 1$ day timescale).
This suggests another energy source such as the jet-powered cocoon,
although this is not definite given
the uncertainties about the observations and
the modelings of the heating and the density profile.
The (prompt) jet can inject energy in Eq.~(\ref{eq:E_MN}) 
that dominates the radioactive energy in Eq.~(\ref{eq:E_r})
for the macronova emission if the opening angle is wide enough in Eq.~(\ref{eq:Ej>Er}).
Then the required ejecta mass is reduced to $M_e < 0.01 M_{\odot}$,
which may be affordable by the conventional dynamical ejection \citep{Hotokezaka+13b}
or the disk outflows with reasonable viscous parameters \citep{Shibata+17,Shibata+17b}.
In addition, the required opacity goes back to a moderate value
for a $t_{\rm MN}\sim 1$ day timescale.

\section{X-ray and radio afterglows of a jet?}\label{sec:afterglow}

The jet interacts with the ISM
and produces afterglow emission by releasing the kinetic energy.
Initially the afterglow emission is beamed into the direction of the jet
and is difficult to detect by off-axis observers.
As the jet is decelerated by the ISM,
the beaming angle becomes wide and
the afterglow begins to be observable by off-axis observers \citep{Sari+99}.
The observable condition is 
\begin{eqnarray}
\frac{1}{\Gamma} \gtrsim \theta_v-\Delta\theta;
\label{eq:offaxisAG}
\end{eqnarray}
i.e., the beaming angle becomes larger than
the viewing angle of the jet edge.
Since the evolution of the Lorentz factor is easily calculated \citep{Sari+99},
we can estimate the rise time of the afterglow from Eq.~(\ref{eq:offaxisAG}) as
\begin{eqnarray}
t_{\rm rise} \sim 14\,{\rm day}
\left(\frac{\theta_v-\Delta\theta}{7^{\circ}}\right)^{8/3}
\left(\frac{E_{\rm iso}(0)/\epsilon_{\gamma}}{3\times 10^{52}\,{\rm erg}}\right)^{1/3}
\left(\frac{n}{10^{-4}\,{\rm cm}^{-3}}\right)^{-1/3},
\label{eq:trise}
\end{eqnarray}
where $n$ is the ambient density (and could be small as discussed below).
For our interest in a wide jet in Eq.~(\ref{eq:Ej>Er}), 
this is usually earlier than the jet break time,
\begin{eqnarray}
t_{\rm jet} \sim 230\,{\rm day}
\left(\frac{\Delta\theta}{20^{\circ}}\right)^{8/3}
\left(\frac{E_{\rm iso}(0)/\epsilon_{\gamma}}{3\times 10^{52}\,{\rm erg}}\right)^{1/3}
\left(\frac{n}{10^{-4}\,{\rm cm}^{-3}}\right)^{-1/3}.
\label{eq:tjet}
\end{eqnarray}
After this time $t_{\rm jet}$,
the Lorentz factor drops below $\Gamma < \Delta\theta^{-1}$ and
the jet's material spreads laterally, 
producing a break in the light curve of the afterglow \citep{Sari+99}.

By using the standard afterglow model, in particular
the spherical model before the jet break \citep{Sari+98},
the characteristic synchrotron frequency and the peak spectral flux
at time $t=15\,{\rm day}\, t_{15{\rm d}}$ are given by
\begin{eqnarray}
\nu_{m} &=& 2.5 \times 10^{7}\,{\rm Hz}\
\epsilon_{B,-6}^{1/2} \epsilon_{e,-1}^{2} E_{52}^{1/2} t_{15{\rm d}}^{-3/2},
\\
F_{\nu,\max} &=& 7.2 \times 10^{3}\, \mu{\rm Jy}\
\epsilon_{B,-6}^{1/2} E_{52} n_{-4}^{1/2} D_{40{\rm Mpc}}^{-2},
\end{eqnarray}
where 
$E=10^{52}$ erg $E_{52}$ is the total energy of the spherical shock,
$n=10^{-4}$ cm$^{-3}\,n_{-4}$ is the ambient density,
$\epsilon_e = 10^{-1} \epsilon_{e,-1}$ and $\epsilon_B = 10^{-6} \epsilon_{B,-6}$
are the energy fractions that go into the electrons and magnetic field, respectively,
$D=40$ Mpc $D_{40{\rm Mpc}}$ is the distance to the source,
and we use the power-law index $p=2.2$ for the accelerated electrons.
Note that $\epsilon_e=10^{-1}$ and $\epsilon_B=10^{-6}$
are within typical values obtained from afterglow observations,
although $\epsilon_B=10^{-6}$ is at the lower end \citep{Kumar+15}.
For typical parameters, the cooling frequency is too high 
and the self-absorption frequency is too low to observe at this time.
The fluxes at radio $\nu=1\,{\rm GHz}\ \nu_{\rm GHz}$ and X-ray 
$\nu=1\,{\rm keV}\ \nu_{\rm keV}$
are estimated as
\begin{eqnarray}
F_{\nu}&=&(\nu/\nu_m)^{-(p-1)/2} F_{\nu,\max}
\nonumber\\
&\sim & 8 \times 10^{2}\, \mu{\rm Jy}\
\epsilon_{B,-6}^{0.8} \epsilon_{e,-1}^{1.2} E_{52}^{1.3}
n_{-4}^{1/2} D_{40{\rm Mpc}}^{-2} \nu_{\rm GHz}^{-0.6} t_{15{\rm d}}^{-0.9},
\label{eq:radio}
\\
\nu F_{\nu} &=& 2 \times 10^{-14}\,{\rm erg}\,{\rm s}^{-1}\,{\rm cm}^{-2}\
\epsilon_{B,-6}^{0.8} \epsilon_{e,-1}^{1.2} E_{52}^{1.3}
n_{-4}^{1/2} D_{40{\rm Mpc}}^{-2} \nu_{\rm keV}^{0.4} t_{15{\rm d}}^{-0.9}.
\label{eq:X}
\end{eqnarray}
The actual fluxes should be less than 
the above spherical estimates by a factor of a few
because we are observing the jet-like edge and 
there is no emission outside the jet-like edge. 

X-ray and radio observations have shown possible counterparts to sGRB 170817A
\citep{GW-EM17,Chandra17,Hallinan+17},
and we can see that they are consistent with 
the above estimates for a typical off-axis afterglow.
First, the rise time in Eq.~(\ref{eq:trise}) fits the observations.
Following early non-detections, delayed X-ray emission is detected $9$ days
after the merger at the position of the macronova by $50$ ks Chandra observations \citep{Chandra17}.
This is followed by the radio discovery $16$ days after the merger
\citep{Hallinan+17}.
To see the allowed parameter region 
for the on-axis isotropic energy $E_{\rm iso}(0)$ 
and opening angle $\Delta\theta$ of the jet,
we plot the line for $t_{\rm rise}=15$ days
in Figs.~\ref{fig:Eiso-dth} and \ref{fig:Eiso-dth2}
({\it magenta curved region})
by varying the density in the range $n=10^{-5}$--$10^{-3}$ cm$^{-3}$.
One reason for adopting these low ISM densities is 
that the host galaxy NGC 4993 is of E/S0 type
(and the other is the faint afterglow fluxes; see below).
As we can see from Figs.~\ref{fig:Eiso-dth} and \ref{fig:Eiso-dth2},
a top-hat jet for sGRB 170817A ({\it black thick line})
can reproduce $t_{\rm rise}=15$ days ({\it magenta curved region})
in the region of typical sGRB parameters ({\it orange square}).
Even if we consider the top-hat jet as an upper limit,
there is a broad parameter space for $t_{\rm rise}=15$ days.

The observed fluxes of the radio and X-ray afterglows are also consistent 
with our estimates in Eqs.~(\ref{eq:radio}) and (\ref{eq:X})
(divided by a few due to the edge effect).
In particular, the flux ratio between radio and X-rays agrees with
the synchrotron spectrum with a typical power-law index
$p \sim 2.2$ for accelerated electrons,
which reinforces the interpretation.
The observed fluxes are not bright, despite the very close distance to the source,
and therefore suggest a low ambient density $n \sim 10^{-3}$--$10^{-6}$ cm$^{-3}$,
not so strange for the E/S0 host galaxy NGC 4993,
unless the jet energy is small $E \ll 10^{51}$--$10^{52}$ erg s$^{-1}$
or the microphysics parameters $\epsilon_e$ and $\epsilon_B$ are small
(see Sect.~\ref{sec:afterglow2} for more discussions).
Both the fluxes are expected to decline similarly in Eqs.~(\ref{eq:radio}) and (\ref{eq:X})
after the peak time, which is later than the rise time $t_{\rm rise}$ 
by a factor of several.
Since the X-rays are now unobservable until early December due to the Sun,
continuous radio observations are important.

\section{Discussions}\label{sec:discuss}

\subsection{sGRB 170817A in other models}\label{eq:prompt}

A top-hat jet is a good approximation if
the energy varies with angle $\theta_v$ more steeply than 
$E_{\rm iso}(\theta_v) \propto (\theta_v - \Delta\theta)^{-4}$ in Eq.~(\ref{eq:Eiso})
outside the opening angle $\Delta\theta$.
If this is not the case,
the jet is structured
(see, e.g., Refs.~\citep[][]{Meszaros+98,Zhang_Meszaros02})
and detectable for a broader range of viewing angles
\citep{Lamb_Kobayashi16,Jin+17,Kathirgamaraju+17,Lazzati+17b}.
Even for the structured jet, the upper limits from a top-hat jet
in Figs.~\ref{fig:Eiso-dth} and \ref{fig:Eiso-dth2} ({\it black thick line})
are applicable.
Although some simulations of the jet propagation
show a structured jet after the breakout (see, e.g., Refs.~\citep[][]{Zhang+03,Morsony+07}),
numerical diffusion of baryons across the jet boundary
is difficult to control under the current resolution \citep{Mizuta_KI13}
and the jet structure down to the observed isotropic energy 
$E_{\rm iso}(\theta_v) \sim 5\times 10^{46}$ is 
difficult to resolve in the present numerical calculations.
Furthermore, the part of the jet that goes to a large viewing angle usually
has a low Lorentz factor $\Gamma \sim \theta_v^{-1} \sim 2 (\theta_v/30^{\circ})^{-1}$
\citep{Gottlieb+17},
and could still be opaque at the observed time $T_{90} \sim 2$ s.
In this case, we expect thermal emission from the cocoon \citep{Ramirez-Ruiz+02,Lazzati+17a}.

The shock breakout of the jet and cocoon from the merger ejecta
could also produce sGRB 170817A (see, e.g., Refs.~\citep[][]{Budnik+10,Nakar_Sari12}).
Although the observations satisfy a relativistic breakout condition
$(T_{90}/2\,{\rm s}) \sim 
(E_{\rm iso}/5\times 10^{46}\,{\rm erg})^{1/2}
(k_B T/160\,{\rm keV})^{-2.68}$ \citep{Nakar_Sari12},
which implies the Lorentz factor of the shock 
$\Gamma \sim k_B T/50\,{\rm keV} \sim 3\, (k_B T/160\,{\rm keV})$,
the required ejecta size at the breakout could be too large 
$\sim c T_{90} \Gamma^2 \sim 5\times 10^{11}$ cm
compared with the fiducial size $\sim c\beta_e t_{\rm br} \sim 10^{10}$ cm.
The large breakout radius could be realized if
the merger ejecta have a faster velocity tail \citep{Kyutoku+14}
than $\sim 0.7 c$ \citep{Kasliwal+17b,Gottlieb+17b}.

The other feasible mechanism is 
the scattering of the prompt emission by the merger ejecta or cocoon
to a large viewing angle
\citep{Eichler_Levinson99,Kisaka+15b}.
In this mechanism,
the scattered $E_{p}$ is similar to the on-axis one,
consistent with the main pulse \citep{Kisaka+17b}.
This is discussed in our other paper \citep{Kisaka+17b}.

\subsection{Macronova in other models}

We should remind ourselves that long-lasting jets following the prompt jet
could also inject less energy
but make a more efficient contribution to the macronova emission
than the prompt jet
\citep{Kisaka+15a,Kisaka+15b,Kisaka+17}.
The longer the injection duration $\sim t_{\rm dur}$,
the smaller the required energy $E_{\rm inj} \sim 10^{48}$ erg $(t_{\rm dur}/10^{4}\,{\rm s})^{-1}$
for the macronova emission
because of lower adiabatic cooling.
Such long-lasting activities are observationally indicated in previous sGRBs:
prompt emission is followed by
extended emission with $t_{\rm dur} \sim 10^{2}$ s and $E_{\rm iso} \sim 10^{51}$ erg
and plateau emission 
with $t_{\rm dur} \sim 10^{4}$ s and $E_{\rm iso} \sim 10^{50}$ erg
(see Refs.~\citep[][]{Barthelmy+05,Rowlinson+13,Gompertz+14,Kisaka+17} and references therein).
The rapid decline of the light curves 
is only produced by activity of the central engine \citep{Ioka+05}.
These long-lasting jets are too faint to observe in sGRB 170817A,
consistent with the observations.
Considering that even the prompt jet is not negligible in this event,
the long-lasting jets could provide
almost all of the energy of the macronova emission, in particular the blue macronova,
without appealing to the radioactive energy.

The red macronova emission at $\sim 10$ day
is an analog of the infrared macronovae observed in 
sGRB 130603B \citep{Tanvir+13,Berger+13}
and 160821B \citep{Troja+16,Kasliwal+17}.
At $\sim 10$ day after the diffusion time in Eq.~(\ref{eq:tdiff}), 
the cocoon is transparent and not relevant to the emission.
The $r$-process radioactivity is widely discussed as an energy source,
and the long timescale is attributed to
the high opacity $\kappa$ 
due to the $r$-process elements, in particular lanthanide elements
\citep{Kasen+13,Tanaka+17}.
However, the required ejecta mass is again relatively huge,
at least $M_e \sim 0.02 M_{\odot}$ for sGRB 130603B \citep{Hotokezaka+13b}
and $M_e \sim 0.03 M_{\odot}$ in this event 
\citep{Tanaka+17b,Utsumi+17,Drout+17,Swift-NuSTAR17,Arcavi+17,Smartt+17,Pian+17,Kasen+17,Kasliwal+17b,Kilpatrick+17,Cowperthwaite+17,Nicholl+17,Chornock+17,McCully+17}.
Similar to the blue macronova at $t_{\rm MN} \sim 1$ day,
which could be powered by a jet,
the red macronova could also imply other energy sources.

One attractive possibility for the red macronova at $\sim 10$ days
is the X-ray-powered model \citep{Kisaka+16}.
This model is motivated by the mysterious X-ray excess 
observed at $\sim 1$--$6$ days with a power-law evolution in sGRB 130603B \citep{Fong+14},
which somehow has a similar flux to the macronova observed in the infrared band.
We can interpret the infrared macronova as the thermal re-emission of
the X-rays that are absorbed by the merger ejecta.
The model naturally explains both the X-ray and infrared excesses observed in sGRB 130603B
with a single energy source such as a central engine like a BH,
and allows for a broader parameter region, 
in particular smaller ejecta mass $\sim 10^{-3}$--$10^{-2} M_{\odot}$
and smaller opacity than the $r$-process model.
The X-ray-powered model is also applicable to 
the macronova at $t_{\rm MN} \sim 10$ day in this event sGRB 170817A \citep{Matsumoto+17}.
Since the X-rays from the central engine are easily absorbed by the ejecta,
it is difficult to find an observational signature of the central engine activity
by off-axis observers in this event, and
it is only possible at late time \citep{Murase+17}.

Note that the comparison between this event and the previous sGRB observations
suggests considerable diversity in the properties of macronovae,
despite the similar physical conditions that are expected in NS--NS mergers
\citep{Fong+17,Gompertz+17}.
While this diversity may come from the merger type (NS--NS vs. BH--NS)
and the binary parameters (mass ratio, spins etc.),
it may imply energy injection from the central engine,
which has more complexities than mass ejection at the mergers.

The blue to red evolution of the macronova is also expected of dust formation
\citep{Takami+14a}.
Dust grains, even a few, provide a large opacity without $r$-process elements
and re-emit photons at infrared wavelengths.
The dust model predicts a spectrum with fewer features than the $r$-process model
and could be tested by spectral observations \citep{Smartt+17,Shappee+17,Pian+17}
(see also Ref.~\citep{Gall+17}).

\subsection{Afterglows in other models}\label{sec:afterglow2}

The X-ray and radio afterglows
may originate from mildly relativistic outflows
rather than the main jet.
Such mildly relativistic outflows could arise from several mechanisms.
First, part of the merger ejecta could be accelerated to a relativistic speed
via the shock breakout at the onset of the merger \citep{Kyutoku+14}.
Second, part of the cocoon material could be mildly relativistic,
depending on the amount of mixing between the jet and the merger ejecta
or the density structure of the merger ejecta
\citep{Nakar_Piran11,Gottlieb+17,Lazzati+17a}.
These mechanisms are difficult to calculate numerically because
only a small part of the mass becomes relativistic
and the relevant range of the density is huge.
Mildly relativistic outflows have a wider opening angle than the main jet,
and therefore a good chance of pointing towards observers.
The outflows are decelerated by collecting $\sim \Gamma^{-1}$ of their rest mass
from the ISM at the time
\begin{eqnarray}
t_{\rm dec} = \frac{1}{4\Gamma^2 c}
\left(\frac{3 E}{4\pi n m_p c^2 \Gamma^2}\right)^{1/3}
\sim 13\,{\rm day}
\left(\frac{E}{10^{49}\,{\rm erg}}\right)^{1/3}
\left(\frac{\Gamma}{2}\right)^{-8/3}
\left(\frac{n}{10^{-3}\,{\rm cm}^{-3}}\right)^{-1/3},
\label{eq:tdec}
\end{eqnarray}
which corresponds to the rise time of the afterglow emission.
This is consistent with the discovery times of the X-ray and radio afterglows.
The expected fluxes in Eqs.~(\ref{eq:radio}) and (\ref{eq:X})
are also consistent with the observations
by choosing appropriate $\epsilon_B$.
Because the energy of the outflows is usually smaller than that of the main jet,
the ambient density
tends to be higher than the jet case 
$n \sim 10^{-3}$--$10^{-6}$ cm$^{-3}$ (see Sect.~\ref{sec:afterglow}).
But too high a density $n > 10^{-2}$ cm$^{-3}$
cannot accommodate the rise time of the afterglows in Eq.~(\ref{eq:tdec})
if the outflow energy is $E < 10^{50}$ erg.

In the above case, we also expect the afterglow emission from the main jet later.
The rise time is months to a year from Eqs.~(\ref{eq:trise}) and (\ref{eq:tjet}),
and the fluxes are potentially 
$\sim 10$--$10^{4}$ times brighter than the initially detected fluxes
from Eqs.~(\ref{eq:radio}) and (\ref{eq:X}).
Therefore, continuous monitoring in radio and X-rays is very important
to reveal the jet and outflows from the NS merger.

The model difference also appears in the image size,
which expands superluminally, depending on the Lorentz factor
$\sim \Gamma c t \sim 8\times 10^{17}\,{\rm cm}\,(\Gamma/10) (t/30\,{\rm d})$
or $\sim 1\,{\rm mas}\,(\Gamma/10) (t/30\,{\rm d})$ at $40$ Mpc.
This might be marginally resolved by VLBI (Very Long Baseline Interferometry)
\citep{Marcote+17} in the case of the off-axis jet in
Eqs.~(\ref{eq:offaxisAG}) and (\ref{eq:trise}).

\subsection{Expected radio flares and X-ray remnants}

The interaction between the merger ejecta and the ISM
produces radio flares \citep{Nakar_Piran11,Piran+13,Hotokezaka+16}
and the associated X-ray remnants \citep{Takami+14b}.
Future observations of these signatures can
reveal the properties of the jet, merger ejecta, and environment,
in particular the ambient density.
As we discuss in Sects.~\ref{sec:afterglow} and \ref{sec:afterglow2},
there remains degeneracy in the ambient density 
from $n < 10^{-5}$ cm$^{-3}$ to $\sim 10^{-2}$ cm$^{-3}$
only with the initial observations of the afterglows.
If $n < 10^{-5}$ cm$^{-3}$, as discussed in Sect.~\ref{sec:afterglow},
the expected radio and X-ray fluxes are very faint 
and difficult to detect even at $D=40$ Mpc \citep{Takami+14b,Hotokezaka+16}.
In addition, the peak time is hopelessly long:
\begin{eqnarray}
t_{\rm dec} = \frac{1}{\beta_e c}\left(\frac{3 M_e}{4\pi n m_p}\right)^{1/3}
\sim 350\ {\rm yr}\ 
\left(\frac{n}{10^{-5}\,{\rm cm}^{-3}}\right)^{-1/3}
\left(\frac{M_e}{10^{-2}M_{\odot}}\right)^{1/3}
\left(\frac{\beta_e}{0.2}\right)^{-1}.
\end{eqnarray}
On the other hand, if the density is moderate $n \sim 10^{-2}$ cm$^{-3}$,
as discussed in Sect.~\ref{sec:afterglow2},
the expected radio and X-ray fluxes are detectable \citep{Takami+14b,Hotokezaka+16}
and the peak time is also within reach.
Therefore, continuous monitoring in radio and X-rays is crucial
for revealing the whole picture.

\subsection{The jet launch time}\label{sec:tj}

The relation $\beta_h \sim 2\beta_e$ between the jet head velocity
and the ejecta velocity at the breakout in Eq.~(\ref{eq:breakout})
is satisfied if $t_{j} \ll t_{\rm br}$.
Otherwise, the situation is similar to 
the breakout from the stellar envelope,
and the cocoon velocity and mass are different 
from Eqs.~(\ref{eq:betaperp}), (\ref{eq:betac}), and (\ref{eq:Mc}).
Since these Eqs.~(\ref{eq:betaperp}), (\ref{eq:betac}), and (\ref{eq:Mc})
are consistent with the observations,
our results imply that
the jet launch time $t_j$ is earlier than $t_{\rm br}< \Delta T_0 \sim 2$ s,
and the delay time $\Delta T_0 \sim 2$ s of the $\gamma$-rays behind the GWs
does not represent the jet launch time.
This information is important for revealing the jet formation mechanism
and could imply that a hypermassive NS formed from two NSs
collapses to a BH earlier than $\sim 2$ s after the NS merger.

\section{Summary}\label{sec:sum}

Prompted by the historical discovery of a binary NS merger
in GW170817 \citep{GW170817},
we calculate EM signals of an associated jet
to reveal its main properties by using multi-wavelength observations.
First, we constrain the isotropic-equivalent energy $E_{\rm iso}(0)$ 
and opening angle $\Delta\theta$ of the jet
by using the $\gamma$-ray observations of sGRB 170817A 
that follows GW170817 after $\sim 1.7$ s \citep{GW-EM17,Fermi/GBM17,Integral17} 
in Sect.~\ref{sec:sGRB}.
We carefully calculate the off-axis emission from a top-hat jet
in Fig.~\ref{fig:Eiso}
to give the most robust upper limits on the $E_{\rm iso}(0)$--$\Delta\theta$ plane
in Figs.~\ref{fig:Eiso-dth} and \ref{fig:Eiso-dth2}
({\it black thick line}).
We again emphasize that the off-axis emission declines more slowly
than the point-source case because of a finite opening angle
in Eq.~(\ref{eq:Eiso}),
which expands the detectable viewing angles.
Second, we examine a possible contribution of the jet energy 
to the macronova emission,
which is blue and very bright at $\sim 1$ day
and difficult to explain by $r$-process radioactivity 
with a canonical ejecta mass $M_e \sim 0.01 M_{\odot}$.
We follow the jet propagation and breakout from the merger ejecta 
by deriving improved analytic descriptions in Sect.~\ref{sec:breakout}.
This gives the injected and released energy from the cocoon
to compare with the radioactive energy 
in Figs.~\ref{fig:Eiso-dth} and \ref{fig:Eiso-dth2}
({\it red vertical line})
and the observed macronova characteristics in Sect.~\ref{sec:MN}.
Third, we calculate the afterglow features of the jet
to obtain the jet and environment properties
from the X-ray and radio observations in 
Figs.~\ref{fig:Eiso-dth} and \ref{fig:Eiso-dth2} 
({\it magenta curved region}) and Sect.~\ref{sec:afterglow}.

Our findings are as follows:
\begin{enumerate}
\item A typical sGRB jet viewed off-axis is consistent with the faint sGRB 170817A.
In particular, a simple top-hat jet can explain sGRB 170817A
with typical isotropic energy $E_{\rm iso}(0) \sim 10^{50}$--$10^{52}$ erg
and a viewing angle in Eq.~(\ref{eq:allow})
as shown in Figs.~\ref{fig:Eiso-dth} and \ref{fig:Eiso-dth2}
({\it black thick line}).

\item The opening angle inferred from sGRB 170817A is also typical
$\Delta\theta \sim 6^{\circ}$--$26^{\circ}$
unless the viewing angle is too large 
$\theta_v > 26^{\circ} + 11^{\circ}(\Gamma/100)^{-1}$
or too small $\theta_v < 6^{\circ} + 3^{\circ}(\Gamma/100)^{-1}$
as shown in Figs.~\ref{fig:Eiso-dth} and \ref{fig:Eiso-dth2}
({\it black thick line}).

\item The jet breakout from the merger ejecta is possible for sGRB 170817A
as shown in Figs.~\ref{fig:Eiso-dth} and \ref{fig:Eiso-dth2}
({\it blue dotted line}).
The breakout time is analytically given in Eqs.~(\ref{eq:tbr1}) and (\ref{eq:tbr2}).

\item The jet-powered cocoon can dominate the blue macronova emission
at $\sim 1$ day, exceeding the radioactive energy,
if the jet opening angle is wide $\Delta\theta > 19^{\circ}$ 
in Eq.~(\ref{eq:Ej>Er}).
This is possible if the viewing angle is 
$20^{\circ} \lesssim \theta_v \lesssim 30^{\circ}$
from Figs.~\ref{fig:Eiso-dth} and \ref{fig:Eiso-dth2}
({\it red vertical line}).
The extra energy from the jet-powered cocoon eases
the requirement of huge mass $M_e \ge 0.02 M_{\odot}$
and small opacity $\kappa \le 0.5$ cm$^2$ g$^{-1}$
for explaining the bright blue macronova at $\sim 1$ day
by $r$-process radioactivity.
If long-lasting jet activity continues after the prompt emission,
which is, however, weak and unobservable,
it could even dominate the macronova emission because of lower adiabatic cooling.

\item 
The jet-powered cocoon has favorable mass in Eq.~(\ref{eq:Mc}) and velocity
in Eqs.~(\ref{eq:betaperp}) and (\ref{eq:betac})
for explaining the timescale $\sim 1$ day
and photospheric velocity $\sim 0.3$--$0.4 c$
of the blue macronova.
According to our improved analytical estimates,
the cocoon velocity and mass fraction do not strongly depend
on the parameters of the jet and merger ejecta.

\item
A typical off-axis jet can reproduce the observed X-ray and radio afterglows
by the standard synchrotron shock model.
The afterglow rise time in Eq.~(\ref{eq:trise}),
determined by the deceleration of the jet and the expansion of the beaming angle,
can match the discovery times $\sim 9$--$16$ days.
The synchrotron fluxes can also fit the observed values.
The faint fluxes despite the nearest sGRB
with the distance $\sim 40$ Mpc observed so far
suggest a low ambient density
$n \sim 10^{-3}$--$10^{-6}$ cm$^{-3}$.

\item 
The X-ray and radio afterglows could instead originate from 
mildly relativistic outflows in the merger ejecta or cocoon.
In this case, the ambient density can be moderate $n \sim 10^{-3}$--$10^{-2}$ cm$^{-3}$,
and brighter afterglows of the main jet could arise months to a year later.

\item
The radio flares and associated X-ray remnants,
caused by the interaction between the merger ejecta and the ISM,
are important for diagnosing in particular the undetermined ambient density.

\item
There is a parameter space for a typical top-hat jet to explain
all the sGRB 170817A, blue macronova, and X-ray and radio afterglows.

\end{enumerate}

A similar GW event with a similar configuration
could occur within $5$--$10$ years.
This is because the merger rate inferred by GW170817
is at the higher end of the previous limits and estimates,
roughly $\sim 2$ event yr$^{-1}$ within $\sim 100$ Mpc
\citep{GW170817},
and the expected typical viewing angle peaks around
$\sim 31^{\circ}$ \citep{Schutz11,Lamb_Kobayashi17}
with the mean $\sim 38^{\circ}$ \citep{Lamb_Kobayashi17}
by considering that GW signals are stronger along the orbital axis.

Given the merger rate and the ejecta mass per merger,
we can see the consistency with the Galactic enrichment rate \citep{Qian00}.
If both the quantities are raised, a tension could appear
in the total abundance, and also in the $r$-process cosmic-ray abundance \citep{Kyutoku_KI16}.
These are interesting future problems.

\subsection{Latest observations}\label{sec:new}

During the refereeing process,
new observations were reported
in radio \citep{Mooley+17},
optical \citep{Lyman+18}, and
X-rays \citep{Ruan+18,Margutti+18,D'Avanzo+18,Troja+18}.
In this final subsection,
we apply our discussions and give possible interpretations.
The observed power-law spectrum
over eight digits of frequency $F_{\nu} \propto \nu^{-0.6}$
suggests synchrotron emission
with the index $p \sim 2.2$ of the electron distribution
where the cooling frequency is above the X-ray band
and the synchrotron frequency is below the radio band.
The light curves show steady brightening $F_{\nu} \propto t^{0.7}$
up to $t \sim 110$ days followed by a possible decline \citep{D'Avanzo+18}.

A simple top-hat jet is not consistent with
the flux rising over one digit in time.
The afterglow of a top-hat jet is thought to rise to the peak
faster than $F_{\nu} \propto t^{0.7}$,
and fall after the peak over a factor of several in time.
Then, if the peak is at $t\sim 10$ days or $t\sim 110$ days,
the late or early flux becomes fainter than the observations, respectively.

We can make a jet model consistent with the observations
by a slight modification of the jet structure.
First, we can easily bring
the rise time of the afterglow in Eq.~(\ref{eq:trise}) to
\begin{equation}
t_{\rm rise} \sim 110\,{\rm days}
\left(\frac{\theta_v-\Delta\theta}{15^{\circ}}\right)^{8/3}
\left(\frac{E_{\rm iso}(0)/\epsilon_{\gamma}}{3\times 10^{52}\,{\rm erg}}\right)^{1/3}
\left(\frac{n}{10^{-4}\,{\rm cm}^{-3}}\right)^{-1/3},
\end{equation}
by using twice the fiducial viewing angle,
$\theta_v-\Delta\theta \sim 15^{\circ}$.
Note that such a viewing angle is consistent with
the off-axis emission model in Eq.~(\ref{eq:allow})
by using a slightly small Lorentz factor
that does not cause the compactness problem
(see Sect.~\ref{sec:spec}).
We can also fit the peak flux by choosing the parameters.
Then we can obtain the early rising $F_{\nu} \propto t^{0.7}$
by introducing the polar jet structure,
which is energetically minor
(see also Ref.~\citep{Lazzati+17c}).
Therefore, the off-axis jet model is currently consistent with the observations
and is not yet excluded.
Note that the jet structure for the afterglow
may not necessarily coincide with the prompt emission structure.

Other models could also explain the steadily rising afterglow.
One possibility is the ambient density structure and/or
the radial jet structure that leads to energy injection at late time.
However, these models generally require a coincidence
between the rising timescale due to these structures
and the rising timescale due to the viewing angle,
and hence are not natural.
Nevertheless, this is one of the few observations of sGRB afterglows beyond $\sim 10$ days
and we cannot exclude these possibilities immediately.
Alternatively, as already argued,
the merger ejecta itself \citep{Kyutoku+13}
or the cocoon \citep{Gottlieb+17b,Lazzati+17a} could produce the afterglow
if these outflows have a relativistic tail.
Further observations are necessary.

\section*{Acknowledgments}

The authors would like to thank
Kazumi Kashiyama, Shota Kisaka, Masaru Shibata, Masaomi Tanaka, and Michitoshi Yoshida for discussions.
This work is partly supported by
``New Developments in Astrophysics through Multi-Messenger Observations
of Gravitational Wave Sources'', No.~24103006 (K.I., T.N.), 
KAKENHI Nos. 26287051, 26247042, 17H01126, 17H06131, 17H06362, 17H06357 (K.I.),
No.~15H02087 (T.N.)
by a Grant-in-Aid from the Ministry of Education, Culture, Sports,
Science and Technology (MEXT) of Japan.

\appendix

\section{Off-axis emission from a top-hat jet}\label{sec:App}

To calculate the off-axis emission from a top-hat jet,
we use the formulation of \citet{KI_Nakamura01}.
A single pulse of sGRBs is well approximated by
instantaneous emission at time $t=t_0$ and radius $r=r_0$
from a uniform thin shell with an opening half-angle $\Delta\theta$
moving radially with a Lorentz factor $\Gamma=1/(1-\beta^2)^{1/2}$.
We assume that the emission is optically thin, and
isotropic in the comoving frame of the jet.
Then we can analytically derive
the spectral flux [erg s$^{-1}$ cm$^{-2}$ eV$^{-1}$]
at the observer time $T$, frequency $\nu$ and viewing angle $\theta_v$ as
\begin{eqnarray}
F_{\nu}(T)=\frac{2 r_0 c A_0}{D^2}
\delta(T)^2 
\Delta \phi(T) 
f\left[\nu/\delta(T)\right],
\label{eq:Fnu_App}
\end{eqnarray}
where $D$ is the luminosity distance, $A_0$ is the normalization,
\begin{eqnarray}
\delta(T) \equiv \frac{1}{\Gamma \left[1-\beta \cos\theta(T)\right]}
\equiv \frac{r_0}{c\beta\Gamma}\frac{1}{T-T_0}
\label{eq:deltaT}
\end{eqnarray}
is a kind of a Doppler factor,\footnote{
The definition of $\delta$ in \citet{KI_Nakamura01}
is the inverse of $\delta$ in our paper.}
and $T_0=t_0-r_0/c\beta$.
The azimuthal angle of the emitting region $\theta(T)$ 
varies from $0$ to $\theta_v+\Delta\theta$ 
for $\theta_v<\Delta\theta$,
and from $\theta_v-\Delta\theta$ to $\theta_v+\Delta\theta$ 
for $\theta_v>\Delta\theta$.
The polar (half-)angle of the emitting region is $\Delta\phi(T)=\pi$
if $\Delta \theta>\theta_v$ and $0<\theta(T)\le\Delta\theta-\theta_v$,
otherwise
$\Delta\phi(T)=\cos^{-1}\left\{[\cos\Delta\theta-\cos\theta(T)\cos\theta_v]/
[\sin\theta_v \sin\theta(T)]\right\}$.

We adopt the broken power-law spectrum in the comoving frame of the jet,
which is similar to the Band spectrum of the observed GRBs \citep{Band+93},
\begin{eqnarray}
f(\nu')=\left(\frac{\nu'}{\nu_0'}\right)^{1+\alpha_B}
\left[1+\left(\frac{\nu'}{\nu_0'}\right)^s\right]^{(\beta_B-\alpha_B)/s},
\end{eqnarray}
where
$\alpha_B$ and $\beta_B$ are the low- and high-energy power-law indexes, respectively,
and $s$ describes the smoothness of the transition.
We adopt $\alpha_B=-1$, $\beta_B=-2.2$ \citep{Preece+00}, and $s=1$ in this paper.
As we integrate the spectrum below,
the choice of the typical frequency $\nu_0'$ does not matter so much
if it is included in the integral range.

The isotropic energy at the viewing angle $\theta_v$ is calculated as
\begin{eqnarray}
E_{\rm iso}(\theta_v)= 4\pi D^2
\int_{T_{\rm start}}^{T_{\rm end}} dT
\int_{\nu_{\min}}^{\nu_{\max}} d\nu\,
F_{\nu}(T),
\label{eq:Eiso_App}
\end{eqnarray}
where 
\begin{eqnarray}
T_{\rm start}&=&T_0+(r_0/c\beta)[1-\beta\cos(\max [0,\theta_v-\Delta \theta])]
\label{eq:Tstart}
\\
T_{\rm end}&=&T_0+(r_0/c\beta)[1-\beta\cos(\theta_v + \Delta \theta)],
\label{eq:Tend}
\end{eqnarray}
and we adopt $\nu_{\min}=10$ keV and $\nu_{\max}=25$ MeV in this paper.
If the sGRB is composed of multiple pulses,
we can add all the isotropic energy.

Approximate scaling of the isotropic energy on the viewing angle 
$E_{\rm iso}(\theta_v)$
is obtained from the above equations
for $\Gamma^{-1} \ll \Delta\theta$.
We can perform the frequency integral first,
\begin{eqnarray}
E_{\rm iso}(\theta_v) \propto 
\int_{T_{\rm start}}^{T_{\rm end}} dT\,
\delta(T)^2 \Delta\phi(T)
\cdot
\delta(T) \int d\nu' f(\nu'),
\label{eq:Eiso_nuint}
\end{eqnarray}
where as long as the spectral peak is included in the frequency integration,
we can approximately regard the last term $\int d\nu' f(\nu')$ as a constant.
For the time integration, we may focus on
the duration $\Delta T$ in which most of the energy is released and perform
$\int dT \to \Delta T$.
Then we can show that the terms in Eq.~(\ref{eq:Eiso_nuint}) scale as follows:
\begin{eqnarray}
\begin{array}{lll}
{\rm For} \quad \theta_v<\Delta\theta, & & \\
\qquad \Delta T \sim r_0/2 c\beta\Gamma^2 = {\rm const.},
& \delta(T) \sim \Gamma, 
& \Delta\phi =\pi,\\
{\rm For} \quad \Delta\theta < \theta_v \lesssim 2 \Delta\theta, & & \\
\qquad \Delta T \sim T_{\rm start}-T_0 \propto \tilde\delta(\theta_v)^{-1},
& \delta(T) \sim \tilde\delta(\theta_v), 
& \Delta\phi \sim \pi,\\
{\rm For} \quad 2 \Delta\theta \lesssim \theta_v, & & \\
\qquad \Delta T \sim T_{\rm end}-T_{\rm start} \propto \delta(\theta_v)^{-1/2},
& \delta(T) \sim \delta(\theta_v), 
& \Delta\phi \sim \Delta\theta/\theta_v \propto \delta(\theta_v)^{1/2},
\end{array}
\label{eq:scale}
\end{eqnarray}
where we define the Doppler factors
\begin{eqnarray}
\tilde \delta (\theta_v) &\equiv& 
\frac{1}{\Gamma [1-\beta\cos(\theta_v-\Delta\theta)]},
\\
\delta (\theta_v) &\equiv& \frac{1}{\Gamma (1-\beta\cos \theta_v)}.
\end{eqnarray}
Note that $\tilde \delta (\theta_v) \sim 2\Gamma/[1+\Gamma^2(\theta_v-\Delta\theta)^2]$
for $\Gamma \gg 1$ and $\theta_v-\Delta\theta \ll 1$.
Note also that, in the above Eq.~(\ref{eq:scale}),
the duration $\Delta T$ in which most energy is released 
is $\Delta T\sim T_{\rm start}-T_0$, not $\Delta T\sim T_{\rm end}-T_{\rm start}$, for 
$\Delta\theta < \theta_v \lesssim 2\Delta\theta$
because the Doppler factor $\delta(T)$ is doubled 
after $\Delta T\sim T_{\rm start}-T_0$ is passed.
Therefore, the scaling of the isotropic energy on the viewing angle is obtained as
\begin{eqnarray}
E_{\rm iso}(\theta_v) &\propto& {\rm const.}
\quad {\rm for} \quad \theta_v<\Delta\theta,
\label{eq:Eiso1}
\\
E_{\rm iso}(\theta_v) &\propto& \tilde \delta(\theta_v)^2
\quad {\rm for} \quad 
\Delta\theta < \theta_v \lesssim 2 \Delta\theta,
\label{eq:Eiso2}
\\
E_{\rm iso}(\theta_v) &\propto& \delta(\theta_v)^3
\quad {\rm for} \quad 
2 \Delta\theta \lesssim \theta_v.
\label{eq:Eiso3}
\end{eqnarray}
Part of the scaling was also derived by Yamazaki et al. \citep{Yamazaki+02,Yamazaki+04}.


\end{document}